\documentclass[a4paper,11pt]{article}
\usepackage{aaskaiid}
\usepackage{xspace}
\usepackage[version=4]{mhchem}
\usepackage{comment}
\usepackage{CJK}
\usepackage{orcidlink}
\newcommand{\arcsec}{\ensuremath{^{\prime\prime}}\xspace}

\title{Unveiling Complex Chemistry in Planet-forming Disks with the SKAO}
\ShortTitle{Chemistry of planet formation with SKAO}

\author[1]{Linda Podio\orcidlink{0000-0003-2733-5372}}
\ShortName{Podio et al.} 

\author[1]{Lisa Giani\orcidlink{0000-0002-7327-9132}}
\author[2]{Catherine Walsh\orcidlink{0000-0001-6078-786X}}
\author[3]{Audrey Coutens\orcidlink{0000-0003-1805-3920}}
\author[4]{Izaskun Jiménez-Serra\orcidlink{0000-0003-4493-8714}}
\author[1]{Claudio Codella\orcidlink{0000-0003-1514-3074}}
\author[5]{Mar{\'i}a Jos{\'e} Maureira\orcidlink{0000-0002-7026-8163}}
\author[6]{Marta De Simone\orcidlink{0000-0001-5659-0140}}
\author[2]{John D. Ilee\orcidlink{0000-0003-1008-1142}}
\author[1]{Manuela Lippi\orcidlink{0000-0001-9185-878X}}
\author[7]{Chin-Fei Lee\orcidlink{0000-0002-3024-5864}}
\author[8]{Romane Le Gal\orcidlink{0000-0003-1837-3772}}
\author[9]{Mayank Narang\orcidlink{0000-0002-0554-1151}}
\author[1]{Giovanni Sabatini\orcidlink{0000-0002-6428-9806}}
\author[1]{Eleonora Bianchi\orcidlink{0000-0001-9249-7082}}
\author[10]{Elenia Pacetti\orcidlink{0000-0003-1096-7656}}
\author[11]{Danai Polychroni\orcidlink{0000-0002-7657-7418}}
\author[12]{Bihan Banerjee\orcidlink{0000-0001-8075-3819}}
\author[5]{Paola Caselli\orcidlink{0000-0003-1481-7911}}
\author[8]{Cecilia Ceccarelli\orcidlink{0000-0001-9664-6292}}
\author[13]{Amin Farhang\orcidlink{0000-0001-7712-697X}}
\author[14]{Antonio Garufi\orcidlink{0000-0002-4266-0643}}
\author[8]{Greta Guidi\orcidlink{0000-0002-7002-8928}}
\author[15]{Adriano Ingallinera\orcidlink{0000-0002-3137-473X}}
\author[16]{Stavro L. Ivanovski\orcidlink{0000-0002-8068-7695}}
\author[17]{Pamela Klaassen\orcidlink{0000-0001-9443-0463}}
\author[8,18]{Ana López-Sepulcre\orcidlink{0000-0002-6729-3640}}
\author[19,20]{Liton Majumdar\orcidlink{0000-0001-7031-8039}}
\author[21,22]{Giulia Perotti\orcidlink{0000-0002-8545-6175}}
\author[5]{Jaime E. Pineda\orcidlink{0000-0002-3972-1978}}
\author[23,8]{Daniel J. Price\orcidlink{0000-0002-4716-4235}}
\author[12]{Manoj Puravankara\orcidlink{0000-0002-3530-304X}}
\author[24]{Pablo Rivière-Marichalar\orcidlink{0000-0003-0969-8137}}
\author[25]{\'Alvaro S\'anchez-Monge\orcidlink{0000-0002-3078-9482}}
\author[10]{Eugenio Schisano\orcidlink{0000-0003-1560-3958}}
\author[16]{Paolo M. Simonetti\orcidlink{0000-0002-7744-5804}}
\author[26]{Leonardo Testi\orcidlink{0000-0003-1859-3070}}
\author[27,6]{Claudia Toci\orcidlink{0000-0002-6958-4986}}
\author[11]{Diego Turrini\orcidlink{0000-0002-1923-7740}}
\author[10]{Alessio Traficante\orcidlink{0000-0003-1665-6402}}
\author[28,29]{Yinhao Wu \begin{CJK*}{UTF8}{gbsn}(吴寅昊)\end{CJK*} \orcidlink{0000-0003-3728-8231}}

\affiliation[1]{INAF, Osservatorio Astrofisico di Arcetri, Largo E. Fermi 5, I-50125, Firenze, Italy}
\emailAdd{linda.podio@inaf.it}

\affiliation[2]{School of Physics and Astronomy, University of Leeds, Leeds, UK, LS2 9JT}

\affiliation[3]{Univ Toulouse, CNES, CNRS, IRAP, Toulouse, France}

\affiliation[4]{Centro de Astrobiolog{\'i}a (CAB), CSIC-INTA, Ctra. de Ajalvir Km. 4, 28850, Torrej{\'o}n de Ardoz, Madrid, Spain}

\affiliation[5]{Center for Astrochemical Studies, Max-Planck-Institut f{\"u}r Extraterrestrische Physik (MPE), Giessenbachstrasse 1, D-85748 Garching, Germany}

\affiliation[6]{ESO, Karl Schwarzchild Srt. 2, 85478 Garching bei M{\"u}nchen, Germany}

\affiliation[7]{Academia Sinica Institute of Astronomy and Astrophysics, Taipei 106216, Taiwan, R.O.C.}

\affiliation[8]{IPAG, Universit{\'e} Grenoble Alpes, CNRS, F-38000 Grenoble, France}

\affiliation[9]{Jet Propulsion Laboratory, California Institute of Technology, 4800 Oak Grove Drive, Pasadena, CA 91109, USA}

\affiliation[10]{INAF, Istituto di Astrofisica e Planetologia Spaziali, Via Fosso del Cavaliere 100, 00133 Roma, Italy}

\affiliation[11]{INAF - Osservatorio Astrofisico di Torino, Strada Osservatorio 20, Pino Torinese, 10020, Italy}

\affiliation[12]{Tata Institute of Fundamental Research, Homi Bhabha Road, Mumbai 400005, India}

\affiliation[13]{School of Astronomy, Institute for Research in Fundamental Sciences, 19395-5531 Tehran, Iran}

\affiliation[14]{INAF, Istituto di Radioastronomia, Via Gobetti 101, I-40129, Bologna, Italy}

\affiliation[15]{INAF, Osservatorio Astrofisico di Catania, via Santa Sofia 78, 95123, Catania, Italy}

\affiliation[16]{INAF, Osservatorio  Astronomico di Trieste, Via Tiepolo 11, I-34143 Trieste, Italy}

\affiliation[17]{United Kingdom Astronomy Technology Centre, Edinburgh, GB, UK}

\affiliation[18]{Institut de Radioastronomie Millim{\'e}trique, 300 rue de la Piscine, Domaine Universitaire, 38406 Saint-Martin d'H{\`e}res, France}

\affiliation[19]{National Institute of Science Education and Research, Jatni 752050, Odisha, India}

\affiliation[20]{Homi Bhabha National Institute, Training School Complex, Anushaktinagar, Mumbai 400094, India}

\affiliation[21]{Niels Bohr Institute, University of Copenhagen, NBB BA2, Jagtvej 155A, 2200 Copenhagen, Denmark}

\affiliation[22]{Max-Planck-Institut f{\"u}r Astronomie, K{\"o}nigstuhl 17, 69117, Heidelberg, Germany}

\affiliation[23]{School of Physics and Astronomy, Monash University, Clayton VIC 3800, Australia}

\affiliation[24]{Observatorio Astron{\'o}mico Nacional (OAN,IGN), Calle Alfonso XII, 3. 28014 Madrid, Spain}

\affiliation[25]{Institut de Ci{\`e}ncies de l'Espai (ICE, CSIC),  Institut d'Estudis Espacials de Catalunya (IEEC), Spain}

\affiliation[26]{Alma Mater Studiorum - Universit{\`a} di Bologna, Dipartimento di Fisica e Astronomia "Augusto Righi", Via Gobetti 93/2, 40129, Bologna, Italy}

\affiliation[27]{Departamento de Fisica aplicada III, ETSI Universidad de Sevilla,  Camino de los Descubrimientos, 41092 Sevilla, Spain}

\affiliation[28]{Shanghai Astronomical Observatory, Chinese Academy of Sciences, Shanghai 200030, People's Republic of China}

\affiliation[29]{School of Physics and Astronomy, University of Leicester, Leicester LE1 7RH, UK}

\abstract{The chemical composition of planets is inherited from that of the natal protoplanetary disk at the time of planet formation. In recent years, we have made huge progress in characterizing disk chemistry. (Sub-)millimeter interferometers, such as ALMA, allowed us to detect emission lines from simple to complex organic molecules and to probe their radial and vertical distribution in disks. On the other hand, JWST has started to unveil the composition of disk ices, and line emission from the innermost disk regions. The advent of SKA will open new domains in the field, by observing emission lines from heavier molecules including heavy carbon chains and rings, and prebiotic molecules with peak emission in the cm range. Moreover, SKA will probe molecular emission from regions which are obscured by dust opacity at mm wavelengths, hence from the disk midplane, and often from the inner $30$ au region. These observations will constrain the initial conditions for disk evolution and planet formation, allowing us to predict the chemical composition of the forming planets and their atmospheres. Comparison with forthcoming results on exoplanet atmospheres and on the chemistry of pristine bodies in the Solar System will provide new hints on the origin and evolution of planetary systems including our own.}

\begin{document}
\maketitle

\section{Introduction}
\label{sect:intro}

Protoplanetary disks are the birthplace of planetary systems. Thus, characterizing the spatial distribution and abundance of molecules in icy and gaseous phases is critical to constrain the chemical inventory that will be incorporated into emerging planets and their atmospheres.
Observationally, the study of the disk chemical properties is hampered by their small sizes (from tens to hundreds of au, i.e. from fractions to a few arcseconds in nearby star forming regions at $d \sim 100-500$ pc), and by molecule freeze-out onto dust grains at temperatures lower than their evaporation temperature. This reduces their gas-phase abundances in the cold midplane and outer disk, below the so-called snow-surfaces. Therefore, characterization of disk chemical structure requires observations that combine high angular resolution ($\le 1"$) and sensitivity (tens of $\mu$Jy).

The advent of ground and space telescopes, such as ALMA at (sub-)millimeter wavelengths, and JWST in the infrared (IR), has allowed huge progresses in the characterization of the chemical composition of  disks.
ALMA large programs such as MAPS \citep{Oberg2021c}, exoALMA \citep{Teague2025}, and AGE-PRO \citep{Zhang2025}, investigated the radial and vertical distribution, and the gas-phase abundances of molecules in protoplanetary disks ($1-10$ Myr), while the FAUST, eDisk, and ALMA-DOT programs surveyed younger Class 0 and I disks (10$^4$-10$^5$ years) \citep{Codella2021,Ohashi2023,Garufi2021}.
Molecules including CO and its isotopologues, simple organics like H$_2$CO, and S-bearing molecules such as CS, SO, and H$_2$CS,  have been observed routinely in the warm molecular layers of Class II planet-forming disks with ages of a few million years \citep[e.g., ][]{Law2021,Legal2021,LeGal2019,Pegues2020}, as well as in their younger counterparts \citep[e.g., ][]{Lin2023,Vanthoff2023,Podio2024,Garufi2022,Garufi2021}, while C- and N-bearing species like c-C$_3$H$_2$, HC$_3$N, CH$_3$CN, have been detected in  carbon-rich disks \citep[e.g., ][]{Ilee2021}.

Dedicated observational efforts searched for the so-called interstellar complex organic molecules (iCOMs), i.e. molecules with more than 6 atoms, which are the seeds of
prebiotic species and precursors of sugars and amino acids \citep[e.g., ][]{Herbst2009,Ceccarelli2023}.
iCOMs have been detected in prestellar cores \citep[e.g., ][]{Bacmann2012} and in hot corinos around protostellar sources \citep[e.g., ][]{Cazaux2003}, and should be inherited by the disk during the collapse phase. However, their detection in disks is hampered by the fact that their sublimation region is confined within the inner few au for Solar-type stars (the typical evaporation temperatures of iCOMs are $\ge 100$ K, e.g.  \citealt{ferrero2020,ferrero2022}), with very low gas-phase abundances due to non-thermal processes in the outer disk \citep{walsh2016}.
Recently, the search for the iCOMs reservoir in disks focused on special classes of disks, such as outbursting disks, where the protostellar luminosity is enhanced by orders of magnitude due to a sudden increase of the accretion activity. This pushes the snowlines outward, releasing the reservoir of iCOMs in gas-phase over a disk region which may extend out to hundred au (see, e.g., the disk V883 Ori,  \citealt{Lee2019}). 
Also some quiescent Class 0 disks are good targets to characterize the disk chemistry, as they are more massive and warmer than their evolved counterpart, which push the snowlines out to $\sim 10$ au \citep[e.g., ][]{vanthoff2020,Maureira2026}, thus allowing in some cases the detection of iCOMs, as in the Class I disk IRAS04302 \citep{Podio2020b}, and the Class 0 disks associated with HH 212 \citep[e.g., ][]{Lee2022} and with the multiple system IRAS 16293-2422 \citep[e.g., ][]{oya2018,maureira2020,zamponi2021}.
Finally, warm transition disks around Herbig Ae stars,  with typical luminosities of 10-100 L$_{\odot}$, show iCOMs emission from the edge of their inner dust cavity which is directly exposed to radiation from the central star, as first noted in the disk of HD 100546 \citep{Booth2021}.

In parallel with the advancements in the detection of gas-phase species obtained with ALMA, observations with JWST in the $0.6-28$~${\mu}\mathrm{m}$ range have allowed the characterization of the ice compositions towards the envelopes of Class 0 and I protostars, as well as the study of the warm gas emission in the innermost disk regions. In the context of the IceAge \citep{McClure2023} and JOYS \citep{vandishoeck2023} programs, JWST has, for the first time, unveiled ice signatures of several iCOMs toward low-mass protostars, ranging from methanol (CH$_3$OH) to CH$_3$CHO, CH$_3$CH$_2$OH, and CH$_3$OCHO \citep[e.g.,][]{Rocha2024,Rocha25,Rayalacheruvu25}, along with the presence of large organic refractories showing broad features corresponding to various complex functional groups \citep[e.g.,][]{Rayalacheruvu25,MCClure2025}. On the other hand, observations of  disks around low-mass stars as part of the MINDS program \citep{Kamp2023} has unveiled bright emission of gas-phase hydrocarbons chains, such as C$_2$H$_2$, di-acetylene (C$_4$H$_2$), and rings, such as benzene (C$_6$H$_6$), which suggests an active hydrocarbon chemistry in the warm inner disk and a C/O ratio larger than 1 \citep[e.g., ][]{Tabone2023}. 

 The main limitation of observations at mm wavelengths is that dust emission in the disk midplane, and up to larger disk height in young disks where dust is not settled, is optically thick hindering our ability to probe the chemical composition of the planet formation region \citep[e.g., ][]{Lee2017a,zamponi2021,Tobin2023,Maureira2026}.
 In addition, the unambiguous identification of iCOMs through their mid-IR bands with JWST remains challenging since they correspond to the different functional groups a molecule has and hence, the observed mid-IR features are produced by multiple molecular species that share the same functional group.
 Finally, observations in the millimetre and IR range cannot probe the emission from complex and heavy carbon chains and rings whose emission at low temperatures ($\le 30$ K) peaks at cm wavelengths.
Recently, radio facilities, such as the Green Bank Telescope (GBT), unveiled complex carbon-bearing molecules towards the starless core TMC-1 in the Taurus molecular cloud \citep[e.g., ][]{Cernicharo2021,McGuire2021,Wenzel2025}, and in the prestellar core L1544 \citep[e.g., ][]{Bianchi2023,giani2025b}. 
Those species could be inherited by the disk at the time of the core collapse, similar to what happens for complex organic molecules and water ice.
This carbon chemistry, however, remains unexplored in disks, because the emission from disks with typical sizes of $100$ au is strongly diluted when observed with single-dish antennas. The JVLA interferometer can reach resolution $\le 1"$, allowing to observe line emission from bright hot corinos located in nearby star-forming regions \citep[e.g., ][]{DeSimone2020} but is not sensitive enough to detect line emission towards disks. 

In this context, the advent of the SKAO will play a pivotal role in investigating the chemical complexity of star-forming regions at cm wavelengths with SKA-Mid  Band 5 receivers ($4.6-15.4$ GHz), from the scales of thousands of au of prestellar cores and protostellar envelopes (Chapter by \citealt{Bianchi01.2026.SKA}), and associated ejection processes (Chapter by \citealt{Sabatini01.2026.SKA}), down to the compact ($\sim 100$~au) scales of planet-forming disks (this chapter). 
In the following, we outline the limitations of current millimeter observations (Sect.~\ref{sect:limits-mm}) and how the SKA Observatory will open a new window on our understanding of disk chemical complexity thanks to the unprecedented combination of angular resolution and sensitivity of SKA-Mid in its final AA4 configuration (197 dishes, and a maximum baseline of 160 km). We discuss the observability of complex organic molecules in protostellar disks (Sect.~\ref{sect:protostellar-disks}), as well as that of complex carbon-bearing molecules in carbon-rich disks (Sect.~\ref{sect:C-molecules}). In Sect. \ref{sect:techniques-for-weak-lines} we present data analysis  techniques to detect weak lines, thus maximizing the exploitation of SKAO observations. The complementarity of SKA observations with existing and ongoing surveys at other wavelengths is outlined in Sect.~\ref{sect:synergies}, while Sect.~\ref{sect:planets} discusses how these observations will constrain models of planetary chemical buildup. This chapter also highlights the crucial need to update chemical networks and to estimate molecular binding energies (BEs) ---through laboratory experiments and theoretical computations--- for a correct interpretation of the observed molecular emission (Sect.~\ref{sect:chem_be}). Finally, Sect.~\ref{sect:comets} shows how comparing the molecular abundances in star and planet-forming systems with those in primitive bodies of the Solar System (SS) help reconstructing the chemical history of the SS. Sect.~\ref{sect:conclusions} briefly summarizes our perspectives to investigate the disk chemistry with the SKA Observatory.



\section{Limits of mm observations and the new radio window with the SKAO}
\label{sect:limits-mm}

Young stellar objects are typically embedded in envelopes of material that gradually dissipate. They are often observed at millimeter and radio wavelengths which penetrate the surrounding cloud. However, in very young cores, dust emission can be optically thick even at millimeter wavelengths. This significantly limits observations, as dust can absorb molecular line emission, hiding the gas behind it. This effect was demonstrated by \cite{DeSimone2020} in the young protobinary system, IRAS 4A, in Perseus. 
At millimeter wavelengths one  protostar of the binary, IRAS 4A2, shows line emission from several iCOMs, while the other component, IRAS 4A1, which is associated to stronger continuum emission, does not \citep{lopezsepulcre2017}.
The authors observed the system at centimeter wavelengths, where the dust emission is optically thinner, and unveiled methanol emission towards both protostars, demostrating that the lack of iCOMs emission when observing with ALMA is due efficient absorption of the dust continuum in the millimetre range. Later, \citet{Frediani2025} showed that even the molecular emission towards IRAS 4A2 was affected by dust opacity in the millimetre range, which biased estimates of both molecular abundances and excitation temperatures.
These studies were pivotal in highlighting the importance of centimeter observations to uncover the true chemical and physical structure of young protostars, which are hidden by dust at shorter wavelengths.

Recent studies show that dust opacity may also affects observations of molecular emission in disks.  We are unable to characterize the molecular composition of the inner disk regions below the dust photosphere, because optically thick dust absorbs the line emission. This effects is more severe in young disks where dust is not settled \citep[e.g., ][]{Lin2021,Michel2022,Villenave2023}, hiding the molecular emission up to larger disk heights. For example, the protostellar outbursting disk V883 Ori  \citep[e.g., ][]{Tobin2023}, as well as the young (Class I, early Class II) disks observed by the ALMA-DOT program show a lack of molecular emission from the inner $30-50$ au region \citep{Garufi2021}. The same is true for some of the  emission lines observed towards the Class II disks targeted by the ALMA LP MAPS \citep{Oberg2021c,Bosman2021}.
The problem of dust opacity also affects the characterization of the molecular composition of the disk midplane in edge-on disks. As an example, ALMA observations of the Class 0 disk HH 212 \citep[e.g., ][]{Lee2017b,Lee2022} and of the Class I disk IRAS04302 \citep[e.g., ][]{Podio2020b,vanthoff2020} allowed recovery of the chemical structures of the respective disks both radially and vertically. However, emission from the disk midplane was absent, not only due to molecule freeze-out but also due to obscuration by optically thick dust.

Another limitation of observations at mm wavelengths is that emission from molecules of increasing complexity and weight, such as long carbon chains and rings, at the low temperatures of the disk midplane ($<30$ K) have their line emission peak at cm wavelengths. 
Recently, two spectral surveys at radio wavelengths of the starless core TMC-1 
have been executed with the Green Bank Telescope (GBT) and the YEBES-40m, namely GOTHAM\footnote{GBT Observations of TMC-1: Hunting Aromatic Molecules} and QUIJOTE \footnote{Q-band Ultrasensitive Inspection Journey to the Obscure TMC-1 Environment}. These ultra-deep surveys discovered new complex carbon-bearing species, such as long cyanopolyynes (up to HC$_{11}$N,  \citealt{2021loomis}), cyclic hydrocarbons such as indene (c-C$_9$H$_8$, \citealt{Cernicharo2021,Burkhardt2021}), and benzonitrile (c-C$_6$H$_5$CN, \citealt{McGuire2021}), and recently the 7-ring PAH Cyanocoronene (C$_{24}$H$_{11}$CN, \citealt{Wenzel2025}).
Follow-up observations with GBT at $8-15$ GHz revealed some of these complex carbon species, such as heavy cyanopolyynes (up to HC$_9$N), in the prestellar core L1544 and protostellar envelope of IRAS 16293 \citep{Bianchi2023,giani2025b}.   
To test if these species are inherited by planet-forming disks with the core collapse requires interferometric observations in this frequency range achieving angular resolution $\lesssim 1"$, to avoid beam dilution of faint line emission from the disk.


Observations with the SKAO will overcome the limitations imposed by current (sub-)millimetre and centimetre observations by: 
(i) probing emission from iCOMs and prebiotic molecules in the central  regions and the  midplane of disks at cm wavelenghts, where the dust is optically thinner \citep{DeSimone2020}  (Sect. \ref{sect:protostellar-disks});
(ii) detecting heavy C-bearing species, such as large cyanopolyynes, whose emission line spectrum shifts towards cm wavelengths at the low temperatures of the midplane (\citealt{Bianchi2023,giani2025b}) (Sect. \ref{sect:C-molecules}).

\section{Chemical complexity in protostellar disks with the SKAO}
\label{sect:protostellar-disks}

\begin{figure}
\begin{center}
\includegraphics[scale=0.75]{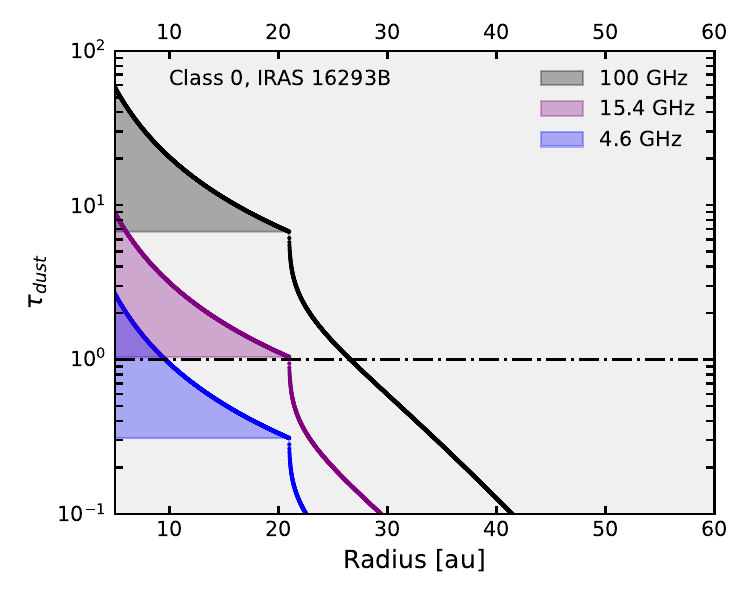}
  \caption{Estimated dust optical depth profiles for the IRAS16293B Class 0 disk at different frequencies. The profile at 100 GHz is estimated using the 1.3-3mm spectral index profile following the methodology in \citet{Maureira2026}. Profiles at 15.4 GHz and 4.6 GHz, frequencies observable with SKAO, are estimated by extrapolating the 100 GHz profile assuming  that the dust emissivity follows a power law with frequency with index $\beta=1$. The shaded region indicates a range of possible values arising from different assumed optical depth in the inner regions, which remains optically thick at 100 GHz and 223 GHz \citep{Maureira2026}. 
  } 
  \label{fig:class0_tau_predictions}
  \end{center}
\end{figure}

Molecular column densities in young disks are larger than measured at the Class II stage (e.g. HH 212, Per-emb-2 compared to the Herbig disks studied by \citealt{Booth2024a,Booth2024b,Booth2025,evans2025}).
However, the chemical composition of the midplane, where planets form, is obscured by optically thick dust. This can lead to the total disappearance of molecular line emission at the location of the dusty disk \citep{DeSimone2020} or to some of the lines to appear in absorption against a hot optically thick structure such as the central part of a young disk \citep{oya2018,zamponi2021}. The optically thick regions observed with ALMA at 1 mm or even 3 mm in young Class 0/I disks can extend to almost the entire disk emitting area \citep{Maureira2026}. Figure~\ref{fig:class0_tau_predictions} shows estimated values of the dust optical depth at 3 mm (100 GHz) as a function of radius for the Class 0 disk IRAS 16293 B with a radius of about 40 au and a gas mass of several 0.1 M$_{\odot}$ \citep{zamponi2021}. The optical depth at 3 mm for the inner region can be larger than 10 and remain above 1 up to $\sim$ 30 au. Class II disks can also become fully optically thick at 3 mm, but this region appears to be more compact than for younger disks \citep[e.g., ][]{Macias2021}. Figure~\ref{fig:class0_tau_predictions} also shows predicted dust optical depth values at frequencies that will be observable with SKAO for IRAS 16293 B. The lower frequencies provided by SKAO will be critical to pierce through the large columns of dust towards this well-known young protostellar disk, allowing us to directly observe the midplane molecular line emission within the 20 au central region.

\subsection{The prototypical protostollar disk HH 212}
\label{sect:hh212}

\begin{figure}
\begin{center}
\includegraphics[scale=0.34]{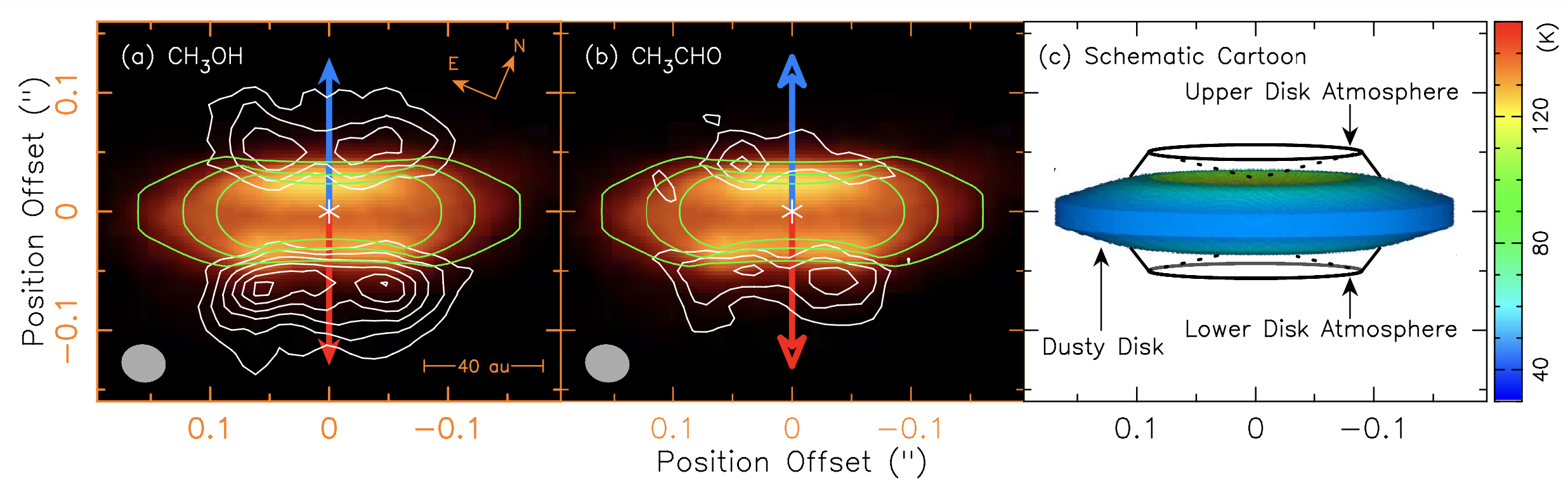}
  \caption{The HH 212 protostellar disk observed with ALMA. The white contours show methanol (CH$_3$OH, left), and acetaldehyde (CH$_3$CHO, middle) emission, while the dust continuum emission is shown in colour scale, with green contours indicating the $\tau =1, 3$ and $5$ surfaces. iCOMs emission is detected in the disk atmosphere, above the $\tau=5$ surface, while the dusty disk midplane is obscured (sketch on the right). Adapted from \citet{Lee2017c,Lee2019}.
  } 
  \label{fig:hh212}
  \end{center}
\end{figure}

An outstanding example of a chemically rich protostellar disk is HH 212, located in the Orion B molecular cloud \citep{Zinnecker1998}. 
HH 212 is associated with a disk wind, and a collimated, rotating, and fast ($\sim$ 100 km s$^{-1}$) jet, which sweeps up a slower and wider molecular outflow \citep[e.g.][and references therein]{Tabone2017,Lee2017c,Codella2019}.
This system has been studied for decades using different telescopes (e.g. IRAM\footnote{https://iram-institute.org/} 30m, and IRAM NOEMA, SMA\footnote{https://lweb.cfa.harvard.edu/sma/}) in typical molecular tracers such as CO isotopologues, SiO, HCO$^+$, and S-bearing species. Recently, ALMA observations at 10~au spatial resolution unveiled the dusty disk oriented nearly edge-on, which extends to a radius of approximately 60 au, and is characterized by an equatorial dark lane due to the high optical depth of the dust emission in the midplane (see Fig. \ref{fig:hh212}, \citealt{Lee2017a}). 

HH 212 is associated with very rich chemistry, with several interstellar complex organic molecules (e.g. CH$_3$OH, HCOOH, CH$_3$CHO, HCOOCH$_3$, NH$_2$CHO) detected using ALMA-Band 6 and Band 7 \citep{Lee2017b,Lee2019,Codella2019}.
\cite{Lee2022} showed that these species have a stratified radial distribution: the outer emission radius increases from 24 au for NH$_2$CHO, to 40 au for CH$_3$OH, and 48 au for H$_2$CO, consistent with their decreasing binding energies to the dust/ice (corresponding to lower sublimation temperatures from the grain mantles). This suggests that the observed iCOMs are thermally desorbed from the grain icy mantles due to the heating by the central protostar. 
The molecule emission, however, is detected only in the disk surface layers, at vertical heights of 40 au from the midplane,  
while no line emission is observed in the midplane due to optically thick dust continuum emission \citep{Lee2019} (see Fig. \ref{fig:hh212}).


The ALMA studies of HH 212 show the level of chemical complexity observable in Class 0 disks, and at the same time the limitations of observations at mm wavelengths. 
Investigating the chemistry of the disk midplane of Class 0 disks requires a combination of high spatial resolution ($\leq$ 100 au) and high sensitivity, 
achieved within a spectral region where dust opacity is not optically thick or sufficiently low to ensure accurate measurement of line intensities.
The search for methanol emission from the midplane of Class 0 disks located in nearby star forming regions present a good case study for SKA in its final AA4 configuration, which will allow to reach angular resolution of $\le 1"$, and the largest collecting area.
The CH$_3$OH-E and A species present several emission lines in SKA-Mid Band 5, corresponding to upper-level energies ($E_{\rm u}$) in the 38-50 K range. When extrapolated to the midplane using the physical conditions derived by \citet{Lee2022} with ALMA for HH 212 ($N_{\rm CH_3OH}$ $\sim$ 10$^{19}$ cm$^{-2}$, n$_{\rm H_2}$ $\sim$ 10$^{9}$ cm$^{-3}$), these lines are expected to be optically thick with brightness temperatures which span from a few to 10 K. Accurate predictions of emission line spectra from protostellar disks and the observability with SKA-Mid Band 5 receveirs is discussed in the following section (Sect. \ref{sect:outbursting-disks}).

\subsection{Search for large organic molecules in outbursting disks}
\label{sect:outbursting-disks}



\begin{figure}[t!]
\begin{center}
\includegraphics[width=15cm]{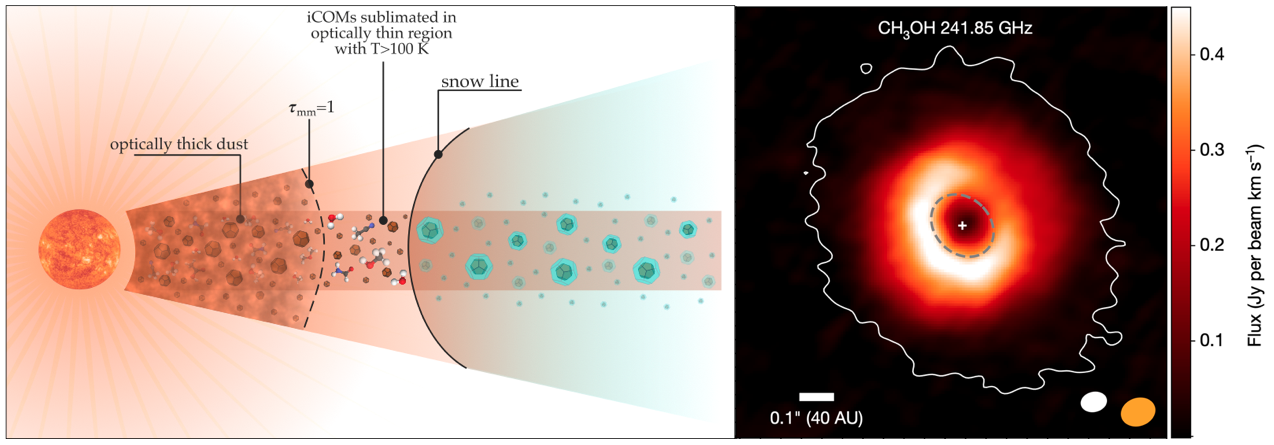}
\caption{{\it Left panel:} Sketch of the V883 Ori outbursting protostar: the luminosity increase of the central source pushes the snowline outwards (from 80 au in the midplane to 160 au in the disk surface layers), releasing iCOMs in gas-phase at dust temperatures $>100$ K. However, the dust continuum optical depth in the inner 40 au screens the emission from complex organics despite their expected high column densities after sublimation \citep[adapted from][]{Leeje2019}.
{\it Right panel}; ALMA observations towards V883 Ori shows a ring of CH$_3$OH emission, because methanol is observed in the region where $T>100$ K (i.e. inside the snowline) and where the dust optical depth becomes lower than 1, 
as shown in the sketch. The lower frequencies covered by the SKAO will allow us to probe iCOMS emission in the innermost 40 au region of these disks thanks to the low dust optical depths at cm wavelengths
\citep[adapted from][]{Tobin2023}. 
}
\label{fig:outbursting_disks}
\end{center}
\end{figure}

The earliest stages of star formation are associated with a rich complex organic chemistry, in particular in the warm inner regions of protostars, with sizes of $\sim 100$ au and temperatures $> 100$ K, also called hot corinos \citep[e.g., ][]{Cazaux2003}. This chemical-rich material should be incorporated in the disks forming in these regions. With time, the disks get colder and colder, and most of the organic molecules deplete on the grains, which makes their detection in the gas phase challenging.
However, some of these sources can undergo episodic accretion events, with an increase in their mass accretion rate of 2 to 5
orders of magnitude
\citep[e.g., ][]{Fischer2023}. 
This leads to luminosity outbursts, during which the bolometric luminosity may increase from a few to hundreds of $L_{\odot}$, and consequently to an increase in the spatial extent of the sublimation region of complex organic molecules (tens to hundred of au), 
enhancing the detectability of their emission lines \citep[see Fig. \ref{fig:outbursting_disks}, ][]{Leeje2019}. 
In the last few years, several disks (V883 Ori, L1551 IRS5, Haro 5A, Haro 5a IRS, V346 Nor, OO Ser, V1057 Cyg) have been found to harbor high abundances of water \citep[e.g.,][]{Cieza2016,Andreu2023,Tobin2023} as well as a high number of complex organic molecules \citep[e.g.,][]{Leeje2019,Bianchi2020,Calahan2024,Cruz2025}, including species such as CH$\rm _3$OH, CH$\rm _3$CHO, CH$\rm _3$OCH$\rm _3$, CH$\rm _3$OCHO, CH$\rm _3$COCH$\rm _3$, C$\rm _2$H$\rm _5$OH, C$\rm _2$H$\rm _5$CN, CH$\rm _3$CN, $\rm ^{13}$CH$\rm _3$CN, $\rm ^{13}$CH$\rm _3$OH, CH$\rm _2$DOH, and HCOOCH$\rm _3$. Despite these achievements, the innermost regions of these disks remain hidden at millimeter/sub-millimeter wavelengths due to the large column densities of dust \citep{Kospal2021}. This can suppress the intensity of the molecular emission in these regions, preventing its imaging toward the innermost $30$ au, the scales relevant for planet formation in planetary systems like our own (see Fig. \ref{fig:outbursting_disks}). In addition, higher column densities of organic compounds are expected toward these highly obscured innermost regions, likely representing the largest organic reservoir in these young disks.

\begin{figure}[h!]
\includegraphics[width=15cm]{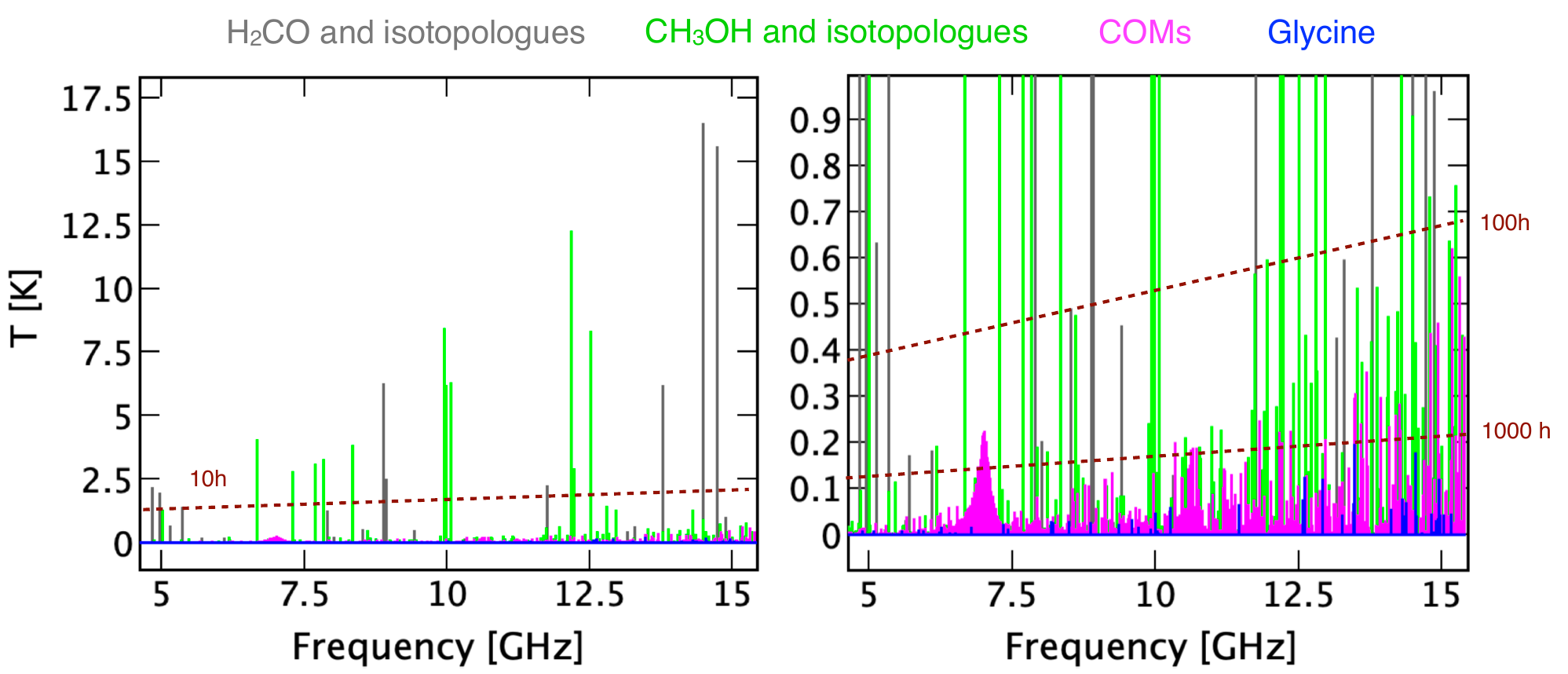}
\caption{Predicted spectra for an outbursting disk with a 0.5\arcsec diameter, obtained with the CASSIS software (CASSIS has been developed by IRAP-UPS/CNRS: \url{https://cassis.irap.omp.eu}). 
The right panel presents a zoom (on the intensity scale) of the left panel. The isotopologues of H$_2$CO are indicated in gray, while the ones of CH$_3$OH are in green. The other iCOMs are shown in pink, and glycine is in blue. The excitation temperature is assumed to be 100 K. The column densities that were used are similar to the ones of the protostar IRAS~16293-2422 \citep{Jorgensen2016,Jorgensen2018,Coutens2016,Lykke2017,Persson2017}. 
The dashed brown line represents 3 times the rms given by the SKA sensitivity calculator by integrating 10, 100, and 1000 hours in configuration AA4 (including MeerKAT antennas), at a spectral resolution of 0.34 km\,s$^{-1}$, and an angular resolution of $\sim$1-2\arcsec, i.e., without spatially resolving the line emission.}
\label{Figure_SKA_outburst_predictions}
\end{figure}

We performed radiative transfer models, assuming local thermodynamic equilibrium (LTE), to determine which molecules could be detected with SKA-Mid in the Band 5 frequency range ($4.6-15.4$ GHz). \citet{Cruz2025} showed that the range of abundances of several iCOMs found in outbursting sources is similar to the more quiescent Class 0 sources. The FUor like sources V883 Ori and L1551 IRS5 also show column densities of CH$_3$OH ($\sim$10$^{19}$ cm$^{-2}$) similar to IRAS~16293-2422 \citep{Yamato2024,Fadul2025a,Jeong2025,Cruz2025}. So we used the column densities of molecules detected towards the Class 0 protostar IRAS~16293-2422 with the ALMA-PILS large spectral survey \citep[e.g.,][]{Jorgensen2016} to obtain the predicted line emission spectrum for a typical outbursting disk. IRAS~16293-2422 is the solar-type protostar with the most complete chemical inventory to date. We assumed an excitation temperature of 100 K, a disk size of 0.5\arcsec and linewidths of 1 km\,s$^{-1}$. The predicted line intensities are shown in Figure \ref{Figure_SKA_outburst_predictions}, along with the $3\sigma$ sensitivity of SKA-Mid in AA4 configuration (including the MeerKAT antennas), for a beam of $1"-2"$, spectral resolution of 0.34 km\,s$^{-1}$, and an observing time of 10, 100, and 1000 hours.  
The sensitivity are obtained using the SKAO sensitivity calculator\footnote{https://sensitivity-calculator.skao.int/ Version 2.2.1.}, and taking into account the additional
MeerKAT antennas equipped with Band 5b receivers, which allows an improvement in sensitivity of a factor $\sim1.4$.

Multiple transitions of CH$_3$OH and H$_2$CO including some of their less abundant isotopologues ($^{13}$C, $^{18}$O, mono-deuterated and doubly-deuterated forms) can be detected in outbursting disks in bands 5a and 5b of SKA-Mid for an observing time of 10 hours.
The detection of larger iCOMs such as dimethylether (CH$_3$OCH$_3$), acetaldehyde (CH$_3$CHO), glycolaldehyde (CH$_2$OHCHO), ethylene glycol ((CH$_2$OH)$_2$), formamide (NH$_2$CHO), methyl formate (HCOOCH$_3$), acetone (CH$_3$COCH$_3$), and ethanol (C$_2$H$_5$OH) requires integration times of 1000 hours. 
Note that this should be considered as a lower limit since the column densities toward the innermost regions of these disks are expected to be higher than observed with ALMA at larger radii, as found for IRAS 16293-2422 or L1551 IRS5 \citep{Jorgensen2016,Cruz2025}. In addition, these regions are highly obscured in the submillimeter range, similarly to what has been found for young protostars such as NGC1333~IRAS4A1 \citep{DeSimone2020}. Disks of FUor-type objects are indeed more massive and smaller in size than regular Class I and Class II disks \citep{Kospal2021}. 
Therefore, SKA has the potential to unveil the chemical complexity toward the innermost regions of disks around outbursting sources, for which their chemistry cannot currently be established with ALMA or NOEMA due to highly optically thick dust. With its very large field of view ($\sim6.7'$ in Band 5), SKA can also simultaneously encompass a large number of sources, helping to identify other targets rich in organic molecules. In Table \ref{tab_outburst_Nmin}, we summarize the column densities needed to detect at least a transition of each species with 3$\sigma$.


Prebiotic molecules such as glycine (the simplest amino acid) could also be searched for in outbursting sources with SKA due to their lines peaking at lower frequencies than smaller iCOMs. 
To detect glycine with SKA AA4 (+MeerKAT) at an angular resolution of $\sim$1-2\arcsec, glycine (conformer I) should have a column density of at least $\sim$ 1.7\,$\times$\,10$^{17}$ cm$^{-2}$. 
At 100 K, it is the conformer I that should have the brightest lines. However, if the excitation temperature is higher ($\sim$ 300\,K), the conformer II would have slightly brighter lines than the conformer I. In that case, the column density needed for detection would be higher ($\sim$6\,$\times$\,10$^{17}$ cm$^{-2}$ for conformer II and $\sim$1\,$\times$\,10$^{18}$ cm$^{-2}$ for conformer I). 
Assuming a water column density of $\sim$2$\times$10$^{20}$ cm$^{-2}$ toward the inner disk region, this translates into a glycine/water abundance ratio of $\sim$0.02\%, which is consistent with those obtained in laboratory experiments of glycine formation in ion and UV-irradiated interstellar ice analogs \citep{munoz-caro2002}, as well as with the values measured in the coma of comet 67P/Churyumov-Gerasimenko  by
the ROSINA mass spectrometer \citep{altwegg2016}. As stressed in \citet{Codella2015} and in the SKA Band 6 White Paper\footnote{\url{https://www.skao.int/sites/default/files/documents/d38-ScienceCase_band6_Feb2020.pdf}}, the detection of glycine or other prebiotic molecules should be facilitated by the expansion of SKA Mid to higher frequencies (up to 25 GHz).

\begin{table}[!h]
\begin{center}
\caption{Minimum column densities needed for the detection of one transition with SKA assuming an excitation temperature of 100 K, a linewidth (FWHM) of 1 km\,s$^{-1}$, a source size of 0.5\arcsec, and 1000 hours of observing time in AA4+MeerKAT configuration with a spatial resolution of $\sim$1-2\arcsec. They are compared to column densities derived towards IRAS~16293B and V883 Ori.}
\label{tab_outburst_Nmin}
\begin{tabular}{ l c c c }
 \hline \hline
 Molecules & $N_{\rm min}$ (cm$^{-2}$) & IRAS~16293B & V883 Ori \\ 
 \hline
 H$_2$CO & 9\,$\times$\,10$^{14}$ & 1.9\,$\times$\,10$^{18}$ & 1\,$\times$\,10$^{18}$\\
 CH$_3$OH & 2\,$\times$\,10$^{16}$ & 1.0\,$\times$\,10$^{19}$ & 4\,$\times$\,10$^{18}$ - 4\,$\times$\,10$^{19}$\\
 CH$_3$CHO & 4\,$\times$\,10$^{16}$ & 1.2\,$\times$\,10$^{17}$ & 6\,$\times$\,10$^{16}$--1\,$\times$\,10$^{18}$ \\
 C$_2$H$_5$OH & 5\,$\times$\,10$^{16}$ & 2.3\,$\times$\,10$^{17}$ & 2.8\,$\times$\,10$^{16}$\\
 CH$_3$OCH$_3$ & 9\,$\times$\,10$^{16}$ & 2.4\,$\times$\,10$^{17}$ & 2\,$\times$\,10$^{16}$--1\,$\times$\,10$^{18}$\\
 HCOOCH$_3$ & 1.2\,$\times$\,10$^{17}$ & 2.6\,$\times$\,10$^{17}$ & 5\,$\times$\,10$^{16}$--1\,$\times$\,10$^{18}$ \\
 CH$_2$OHCHO & 8\,$\times$\,10$^{15}$ & 6.8\,$\times$\,10$^{16}$ & - \\
 aGg-(CH$_2$OH)$_2$ & 1.3\,$\times$\,10$^{16}$ & 1.1\,$\times$\,10$^{17}$ & - \\
 gGg-(CH$_2$OH)$_2$ & 2\,$\times$\,10$^{16}$ & 1.0\,$\times$\,10$^{17}$ & 3.6\,$\times$\,10$^{16}$ \\
 CH$_3$COCH$_3$ & 1.3\,$\times$\,10$^{16}$ & 1.7\,$\times$\,10$^{16}$ & 1.7--4.4\,$\times$\,10$^{16}$ \\
 NH$_2$CHO & 5\,$\times$\,10$^{15}$ & 1.0\,$\times$\,10$^{16}$ & 6\,$\times$\,10$^{14}$ \\
 HOCH$_2$CN & 4\,$\times$\,10$^{16}$ & 3.0\,$\times$\,10$^{15}$ & 3.4\,$\times$\,10$^{16}$ \\
NH$_2$CH$_2$COOH & 1.7\,$\times$\,10$^{17}$ & - & -\\
\hline
\end{tabular}
\end{center}
{\bf Notes:} References for the column densities in IRAS~16293B: \citet{Jorgensen2016,Jorgensen2018,Coutens2016,Lykke2017,Persson2017,Ligterink2021}.
References for the column densities in V883 Ori: \citet{Yamato2024,Fadul2025b,Fadul2025a,Jeong2025}.
\end{table}





\section{Complex C-bearing species in protoplanetary disks with the SKAO}
\label{sect:C-molecules}



\subsection{From prestellar cores to disks}


\begin{figure}
\begin{center}
\includegraphics[width=\textwidth]{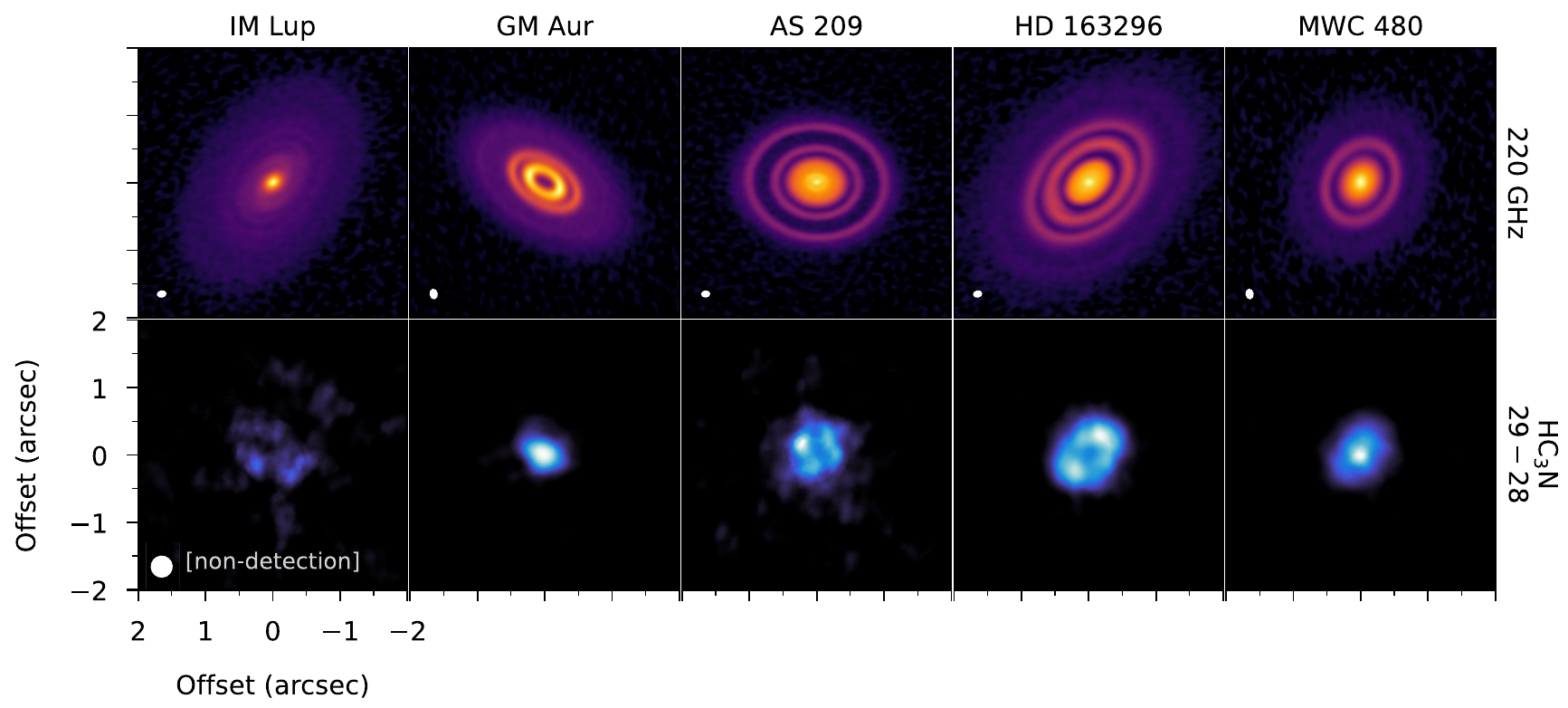}
  \caption{Continuum maps at 220 GHz (top) and integrated intensity (moment 0) maps of HC$_3$N (bottom) in the disks observed by the ALMA Large Program MAPS \citep{Ilee2021}. The ellipses indicate the beam sizes (identical for all integrated intensity maps). The maps are normalized and the intensity is in logarithmic and linear scales for the continuum and integrated intensity (or moment 0) maps, respectively.} 
  \label{fig:maps-disks}
  \end{center}
\end{figure}

More evolved Class II disks (age of a few Myr) observed by the ALMA Large Program MAPS \citep{Oberg2021c} show bright CH$_3$CN and HC$_3$N emission.
Typical disk-averaged column densities and excitation temperatures derived from rotational diagrams are N(HC$_3$N) = $2-8 \times 10^{13}$ cm$^{-2}$, $T_{\rm ex} = 20-50$ K  (see Fig. \ref{fig:maps-disks}, \citealt{Ilee2021}).
While HC$_3$N is the starting point for the formation of longer carbon chains, such as cyanopolyynes (HC$_{2n+1}$N, with $n\ge2$), their detection requires observations at cm wavelengths coupling high sensitivity and high-angular resolution, to avoid the dilution of the faint line emission from the disk.

\begin{figure}
\begin{center}
\includegraphics[scale=0.45]{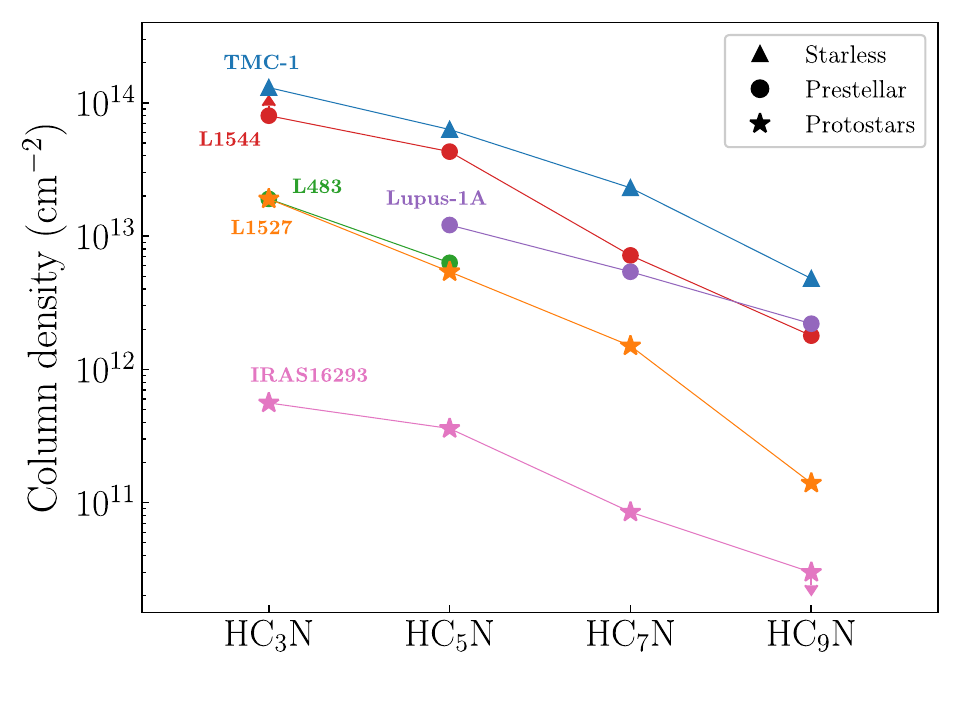}
\includegraphics[scale=0.46]{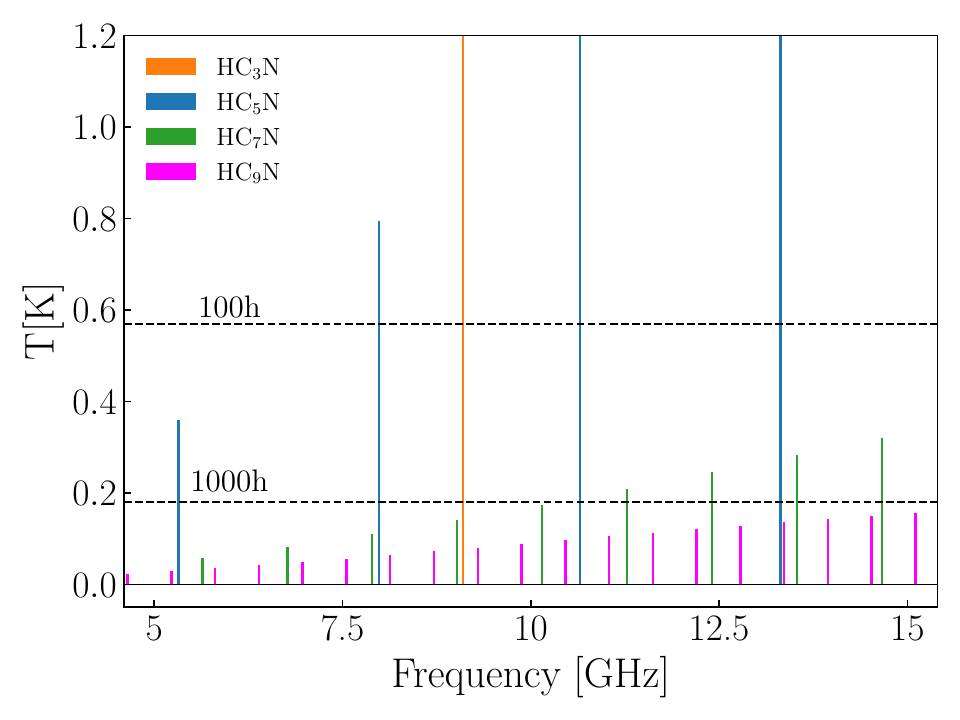}
\caption{
{\it Left panel:} Column densities of cyanopolyynes, HC$_{\rm 2n+1}$N,
in the dark cloud TMC-1, in the prestellar cores L1544, Lupus-1A, L483, in the protostars L1527 and IRAS 16293. Column densities are from: \citet{Bianchi2023,giers2023,giani2025b} (L1544); \citet{Jaber2014,Lindberg2016,giani2025b} (IRAS16293); \citet[][and references therein]{Oyama2020} for the other sources.
{\it Right panel:} Predicted line intensities of HC$_3$N and cyanopolyynes which are still undetected in disks (HC$_5$N, HC$_7$N, HC$_9$N). The predictions are obtained assuming LTE at $T=20$ K, FWHM=1 km\,s$^{-1}$, and filling factor $ff=1$. We assume that the column density of HC$_3$N is $2 \times 10^{14}$ cm$^{-2}$ (a factor 2 larger than estimated by  \citealt{Ilee2021}, taking into account the  attenuation of line emission due to dust opacity at mm wavelengths), and that the column densities of cyanopolyynes scale accordingly to what observed in the prestellar core L1544 (left panel, \citet{giani2025b}). The horizontal dashed lines indicate the 3$\sigma$ sensitivity obtained with SKA-Mid AA4 at a spectral resolution of 0.34 km\,s$^{-1}$, and for a beam of $1.39" \times 1.33"$ in 100, and 1000-hour integration times (from the SKAO sensitivity calculator).
}
  \label{fig:giani2025}
  \end{center}
\end{figure}

To understand the feasibility of the search of these species in disks with SKA-Mid, we refer to recent GBT observations in the X ($8.0-11.6$ GHz) and Ku band ($13.5-15.4$ GHz) of polyynyl radicals, C$_{\rm 2n}$H, and cyanopolyynes, HC$_{\rm 2n+1}$N, in the prototypical prestellar core L1544 (\citealt{Bianchi2023,giani2025b}).
The column densities of C$_{\rm 2n}$H and  HC$_{\rm 2n+1}$N inferred for L1544 exponentially decay with the number of carbon atoms in the molecule, with a shallower trend for cyanopolyynes.
Interestingly, the same exponential decay is observed in other prestellar cores, in dark clouds, and in protostellar sources (see Fig.~\ref{fig:giani2025} from \citealt{giani2025b} and \citealt{Oyama2020}).
This supports a chemical connection between the C$_{\rm 2n}$H and HC$_{\rm 2n+1}$N families, and suggests inheritance along the star formation process and that the same gas-phase chemistry is at work 
at all stages and in different star-forming regions.

To verify the observability of heavier cyanopolyynes from disks, we assume that these inherit their chemical composition from the prestellar stage, and hence that the column densities of heavier cyanopolyynes HC$_{\rm 2n+1}$N scale with the same trend observed in L1544 with respect to the N(HC$_3$N) inferred from the ALMA-MAPS observations for the disks shown in Fig.~\ref{fig:maps-disks}.
Hence, the disk-integrated line intensities in the SKA-Mid Band 5a and 5b frequency range ($4.6-15.4$ GHz) are predicted assuming LTE at T=20 K, column densities of $10^{14}$ cm$^{-2}$ (HC$_5$N), $2\times10^{13}$ cm$^{-2}$ (HC$_7$N), and $6\times10^{12}$ cm$^{-2}$ (HC$_9$N), line FWHM$=1$ km\,s$^{-1}$, and filling factor ff=1 (no beam dilution).
The right panel of Fig. \ref{fig:giani2025} shows the predicted line intensities in the SKA-Mid Band 5 frequency range, and the $3\sigma$ sensitivity obtained in AA4 configuration (including MeerKAT antennas) for a beam of $1.39" \times 1.33"$ and spectral resolution of 0.34 km\,s$^{-1}$, in 100 and 1000-hour integration times, obtained using the SKAO sensitivity calculator.
The figure shows that cyanopolyynes emission from disks can be observed only by using the full array configuration to achieve  high sensitivity at a resolution of $\sim1.3\arcsec$, avoiding beam dilution for a typical disk size of $100-200$ au in a nearby star-forming region (see Fig. \ref{fig:maps-disks}). In particular, HC$_3$N and HC$_5$N emission can be detected after 100 hours in AA4, while 1000-hour integration time is needed to detect fainter emission from heavier cyanopolyynes (HC$_7$N) for the assumed column densities.


\subsection{Emission from carbon-rich disks}


\begin{figure}
\begin{center}
\includegraphics[width=\textwidth]{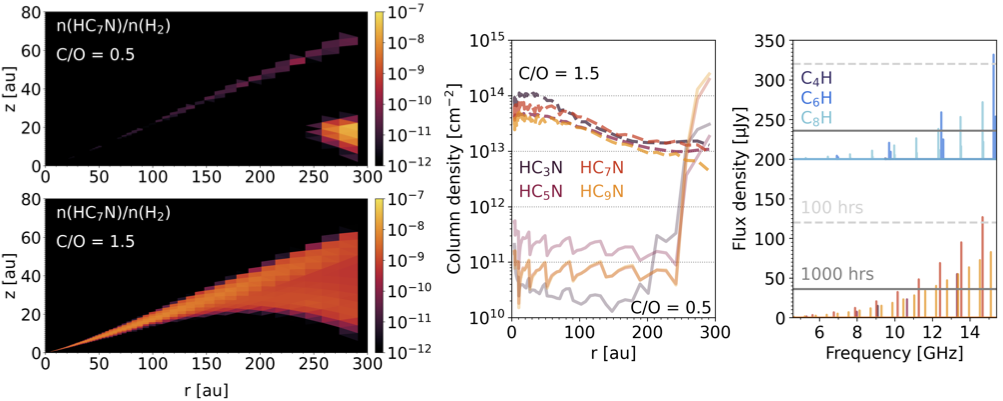}
  \caption{{\it Left panels:} Modelled fractional abundance (with respect to \ce{H2}) of \ce{HC7N} as a function of radius, $r$, and height, $z$, for a model with C/O = 0.5 (top), and one with C/O = 1.5 (bottom).
  {\it Middle panel:} Vertically-integrated column density (cm$^{-2}$) of cyanopolyynes ($\mathrm{HC_{n}N}$) as a function of radius, $r$, for the model with C/O = 0.5 (solid lines), and one with C/O = 1.5 (dashed lines).
  {\it Right panel:} Disk-integrated line spectra at SKA-Mid Band~5 frequencies for cyanopolyynes (colour scheme is the same as for the middle plots) and $\mathrm{C_{n}H}$ species for the model with C/O = 1.5. The horizontal lines indicate the expected sensitivity reached for 100 hours (light-grey dashed line) and 1000 hours (dark-grey solid line) of observation time with SKA-Mid in AA4 configuration (beam of $\sim1.3"$ and spectral resolution of 0.34 km\,s$^{-1}$).
}
  \label{fig:hydrocarbonsSKA}
  \end{center}
\end{figure}

Observations of the outer regions of protoplanetary disks with ALMA suggest that many are carbon-rich with C/O$~> 1$.  
This is evidenced in the detection of strong emission from small hydrocarbons such as \ce{C2H} for which the line emission can be as bright as that from $^{13}$CO \citep[e.g.,][]{Bergin2016,Miotello2019,Bosman2021b}, as well as by observations of CS \citep{Legal2021}. 
This excess of carbon can drive the formation of longer-chain hydrocarbons which have their lower-energy rotational lines at cm wavelengths (such as those discussed above).

To illustrate this, in Fig.~\ref{fig:hydrocarbonsSKA} we show the abundance of \ce{HC7N} as a function of radius, $r$, and height, $z$, calculated using the gas-grain model described in \citet{Walsh2015} and \citet{Ilee2026} for a disk around a generic T~Tauri star (calculated as described in \citealt{Bruderer2012}). 
The top-left panel shows the results for a model with C/O = 0.5 (i.e., similar to solar) and the bottom-left that for a model with C/O = 1.5 and depletion of oxygen by a factor of 100.  
In the carbon-rich models, the abundance of \ce{HC7N} increases significantly in the warm molecular layer of the disk by around a factor of 100 to 1000.  
The excess carbon available enables the efficient building of longer-chain hydrocarbons through ion-molecule chemistry. 
This leads to a large jump in the vertically integrated column density within 200 au from $\sim 10^{11}$~cm$^{-2}$ to $\sim 10^{13} - 10^{14}$~cm$^{-2}$ (middle panel of Fig.~\ref{fig:hydrocarbonsSKA}). 
The behaviour of the other members of the cyanopolyynes is similar with the carbon-rich models showing large increases in column density across the disk when compared with the oxygen-rich model.
In the oxygen-rich models (C/O = 0.5), the increase in the abundance (and thus column density) of \ce{HC7N} is seen in the outer disk midplane ($> 250$~au) due to the efficient conversion of CO and other oxygen-bearing molecules to large complex organic molecules via grain-surface chemistry, creating a carbon-rich gas which develops naturally in the chemistry. 

The right panel of Fig.~\ref{fig:hydrocarbonsSKA} shows the disk-integrated spectra assuming LTE across the frequency range of interest for SKA-Mid for the carbon-rich model.  
The simulated spectra for the $\mathrm{C_nH}$ species are offset by 200~$\mu$Jy for clarity. 
The larger species in particular (\ce{HC7N}, \ce{HC9N}, \ce{C6H}, and \ce{C8H}) reach appreciable flux densities ($\gtrsim 50~\mu$Jy) at the higher frequencies of SKA ($> 11$~GHz). 
These lines may be observable with SKA-Mid in AA4 configuration (beam of $1.39" \times 1.33"$ and spectral resolution of 0.34 km\,s$^{-1}$) but only for integration times on the order of 1000~hours. 

\section{Improving the efficiency of detecting weak lines}
\label{sect:techniques-for-weak-lines}

As the previous sections have described, many of the transitions of (complex) molecules are weak even across the highest frequency ranges that will be offered by SKA Mid (Band 5b, 8--15 GHz).  While this can be alleviated by moving to even higher frequencies, e.g.\ the proposed Band 6 (15--25 or even 50 GHz; see SKA Memo 20-01\footnote{https://www.skao.int/sites/default/files/documents/d38-ScienceCase\_band6\_Feb2020.pdf}) there are a variety of techniques that can be applied to observations of molecular lines in disks to improve the efficiency of detection.

\subsection{Image plane techniques}

The most accessible of these methods involves stacking of individual transitions to improve the effective per channel sensitivity.  Since many of the complex molecules of interest have many tens (or even hundreds) of individual rotational transitions across the large frequency range that will be covered by the spectral windows of SKA-Mid, this can result in a significant increase in the effective signal-to-noise ratio (SNR) which scales with the square root of the number of targeted transitions.  This technique is used widely in studies of line emission across multiple astrophysical objects, for example it has been employed to obtain the first detection of methanol, CH$_3$OH, in a protoplanetary disk \citep{walsh2016}. This first detection of CH$_3$OH, and hence the efficacy of the stacking methodology, has been recently validated with the detection of the individual CH$_3$OH lines in the disk of TW Hya by \citet{Ilee2026}.

\begin{figure}
    \centering
    \includegraphics[width=\linewidth]{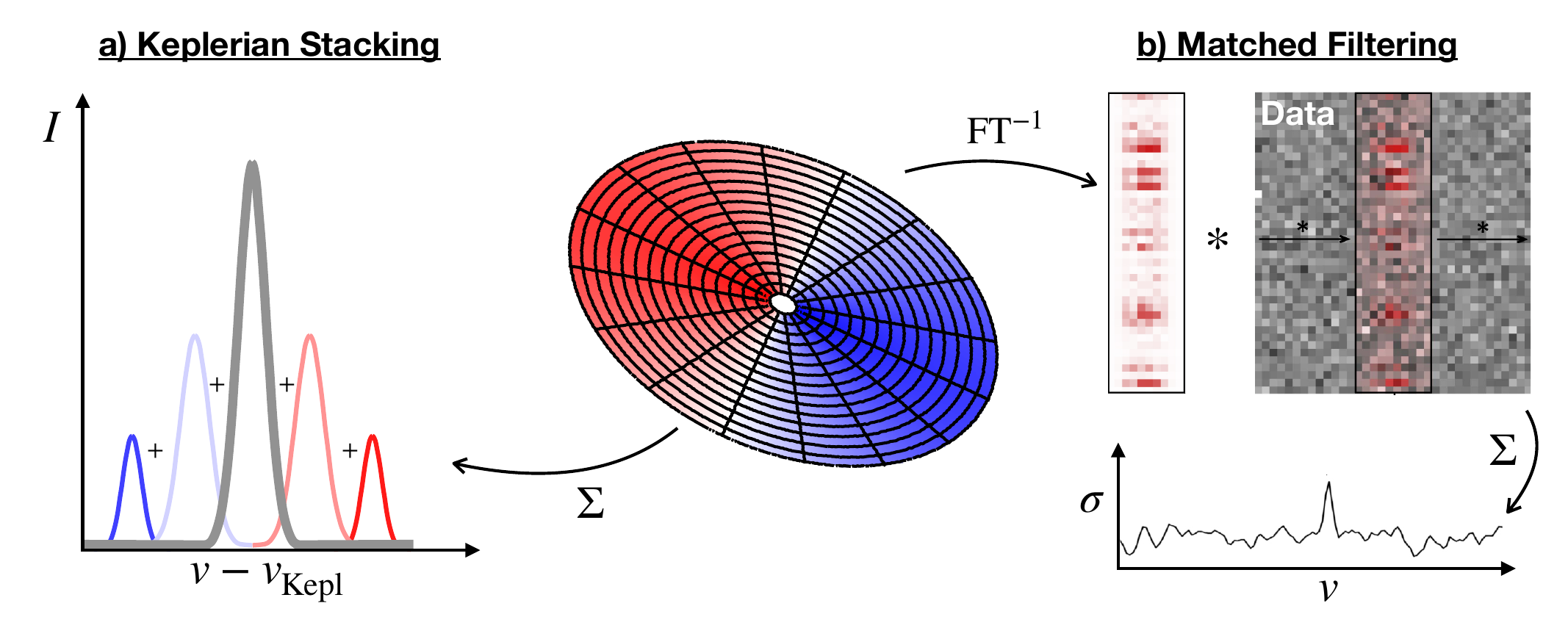}
    \caption{Diagram depicting various methods to improve the signal-to-noise of line observations from Keplerian disks: (a) the Keplerian stacking method \citep[e.g.][]{Yen2016}, and (b) application of a matched filter \citep[based on][]{Loomis2018}.}
    \label{fig:stacking_and_filtering}
\end{figure}

The stacking method can be further improved by exploiting the predictable morphology of line emission in a protoplanetary disk in position-position-velocity (PPV) space.  The characteristic 'butterfly' pattern of emission in strong lines such as $^{12}$CO (e.g. \citealt{Horne1986,Rosenfeld2013}) provides a predictable template over which to search for weaker line emission.  Using known properties of the disk, namely the inclination and position angle of the disk and the distance and mass of the central star, emission can be extracted over a specified radial and azimuthal range.  This emission is summed together after correcting for the Doppler shift caused by Keplerian rotation, yielding averaged or integrated spectra on a common centroid velocity reference frame (see Figure \ref{fig:stacking_and_filtering}a)\footnote{see \citealt{Yen2016} for the first application of this technique and \citealt{Teague2019} for a widely used tool to apply it.}.  The achievable increase in SNR using this 'Keplerian stacking' method is a function of the disk orientation and emission morphology, but typical increases of factors of 5-10 are not uncommon \citep{Bergin2024}.  Combining signals in this way can also be exploited to provide super-spectral resolution.  Here the stacked line profile can reveal underlying spectral features well below the native spectral resolution of the original observations (such as hyperfine splitting, see e.g. CN in TW Hya by \citealt{TeagueLoomis2020}).

\subsection{Visibility plane techniques}

A fundamental requirement of the above methods is the imaging process, which involves trade-offs between angular resolution and sensitivity. Imaging also has a non-trivial computational cost for high resolution observations with large bandwidths such as those that will be provided by SKA Mid.  
An alternative approach to weak line detection involves searching for the predictable spatio-kinematic morphology of molecular line emission from Keplerian disks in the visibility data itself.  \citet{Loomis2018} outline a technique to do this involving matched filtering, which is in essence a cross-correlation of the expected emission pattern (the filter kernel) swept across the full spectral window of an observation.  This yields an impulse response spectrum which gives a measure of how well matched the kernel is to the underlying observations (see Figure \ref{fig:stacking_and_filtering}b).  The matched filtering technique has been demonstrated to lead to a per channel sensitivity increase of up to 500\% \citep{Carney2017}, which would be the equivalent of a factor of 25 increase in integration time.  The technique itself is also fast, enabling a robust identification of all lines across a broad spectral range in a small fraction of the time it would take to image the same dataset \citep{Loomis2020}.  These advantages have seen the technique widely used in studies of weak lines from protoplanetary disks with ALMA \citep[see, e.g.][]{Carney2019, Ilee2021, Booth2024a, Booth2024b, Booth2025, Ilee2026}.
While our above examples demonstrate the application of these techniques to planet-forming disks, they can be applied to {\it any} astronomical source with a well defined structure in position and velocity space.  This highlights the importance of ensuring access to calibrated visibilities in order to take full advantage of SKA Mid's transformative abilities.

\section{The synergy of SKA observations with ongoing and future programmes aimed to characterize the disk and exoplanets chemistry}
\label{sect:synergies}


Observations of the disk chemical complexity obtained with the SKA-Mid in Band 5, will be complementary to ongoing and future programmes surveying the chemistry of planet-forming disks in other frequency ranges, as well as to programs dedicated to characterize the chemical composition of exoplanets. In this section we briefly review the complementarity with some of these programs.

{\bf Observations of the disk chemical composition at mm wavelengths}
A number of ALMA programs explored the disk chemistry, such as, e.g., FAUST \citep{Codella2021} and eDisk \citep{Ohashi2023} dedicated to protostellar disks (Class 0 and I), and MAPS \citep{Oberg2021c}, and AGE-PRO \citep{Zhang2025}, dedicated to Class II disks of a few Myr.
A number of other programs are still in progress, and the results and observations will be available soon. 
As an example, the ALMA large program COMPASS (J{\"o}rgensen et al. in prep.) investigates the chemical composition of 11 solar-type protostars including outbursting sources such as V883 Ori. Their chemical content will be determined through ALMA large spectral surveys in Band 7. By probing longer carbon chains at cm wavelengths, and complex organic molecules in the regions obscured by dust opacity in the mm, SKA will provide a complementary view of the sources observed by COMPASS, allowing, eg, to evaluate the attenuation due to dust opacity and correct the abundances obtained from ALMA observations, while simultaenously delivering dections or upper limits on the complex carbon species not observable with ALMA. This will  lead to a more complete knowledge of the chemistry of protostellar disks. 

Edge-on disks also offer a unique opportunity to directly investigate the disk vertical molecular stratification, and thermal structure, providing important insights into planet formation. This has been recently demonstrated by the ALMA observing program \texttt{EODS} \citep{Guilloteau25, Dutrey2025, Foucher2025}.
The ongoing {\it DiskStrat} ALMA Large Program (Le Gal et al. in prep.) is specifically designed to map the vertical chemical stratification of protoplanetary disks, targeting nine edge-on systems with high-resolution observation ($\sim0.15$'') of key molecular tracers such as CO isotopologues, small organics (e.g.\ce{H2CO}), cyanides, ions (e.g. \ce{HCO+}), small hydrocarbons, and S-bearing species. By resolving the vertical distribution of gas and dust, and leveraging JWST complementary data for ice composition, DiskStrat will constrain the interplay between dust settling, UV penetration, and molecular freeze-out. The SKAO's complementary cm-wavelength observations will extend this work by probing deeper into the disk accessing colder and denser midplane regions, where dust opacity at mm wavelengths obscures molecular emission, and by detecting heavier carbon chains and prebiotic molecules that are invisible to ALMA. Follow-up SKA observations of the DiskStrat sample will thus deliver a fully 3D multi-phase and multi-wavelength picture of disk chemistry, from the UV-irradiated surface to the shielded, cold planet-forming midplane, enabling a robust link between disk composition and the initial conditions for planet formation.

{\bf Observations of the disk chemical composition at IR wavelengths}
On the other hand, JWST presents an unprecedented opportunity to investigate the molecular inventory in both the gas and solid (ice) phases within protostellar envelopes and disks, e.g. in the context of programs such as Ice Age \citep{McClure2023} and JOYS \citep{vanDishoeck25}, as well as in $1-10$ Myr protoplanetary disks targeted by the MINDS  \citep{Henning24,Kamp2023} and JDISC \citep{Arulanantham25,Pontoppidan24} programs.  Owing to its exceptional sensitivity and broad spectral coverage, particularly in the mid-IR, JWST enables the detection and characterization of a wide range of molecular features, including key volatile species such as H$_2$O, CO$_2$, CH$_4$, HCN, C$_2$H$_2$, and several hydrocarbons, including C$_6$H$_6$, and numerous complex organic molecules, directly in their ice form \citep{Nazari24, Rayalacheruvu25, Tyagi25, Rocha25,Rocha2024}.  
As protostellar envelopes are heated by the central source or through shocks, the ices sublimate from dust grains and enter the gas phase \citep[e.g.,][]{Chen_Y24,Perotti2020,Noble2017}. Once in the gas phase, these molecules become accessible to observations with ALMA and SKA, which are well-equipped to detect low-energy rotational transitions of cold gas. This provides a complementary view to JWST's infrared capabilities, which are more sensitive to warmer gas associated with the inner disk regions \citep{Arabhavi24,Tabone2023,Perotti2023} or with shocks occurring along  disk winds and outflows
\citep{Rubinstein24,Delabrosse24,Salyk24}.

JWST programs such as JDISC and MINDS revealed that protoplanetary disks are rich in molecular emission, ranging from simple species such as H$_2$, CO, CO$_2$, OH, and H$_2$O to hydrocarbons \citep[e.g.,][]{Arulanantham25,Arabhavi24,Kanwar24,Patapis25,Tabone2023,Perotti2023}.
Understanding the formation and transport pathways of molecules from the protostellar envelopes to disks requires a synergy between JWST, ALMA, and the SKA. Recent JWST MIRI MRS observations provide compelling evidence that disk substructures play a key role in regulating the inward drift of icy pebbles toward the inner disk regions \citep[e.g., ][]{Gasman2025,banzatti2023}. In a sample of 21 disks, \cite{Krijt25} finds that the 1500~K~/~6000~K H$_2$O line flux ratio measured with JWST can be a tracer of cold water vapor and pebble delivery across the snowline, with the ratio correlating with the location of the innermost dust gap observed in ALMA continuum data. With SKA we should be able to map the location of these large pebble sized grains in disk and investigate if the location of these large grains also correlates with the H$_2$O flux ratio.   

{\bf Observations of the exoplanets chemical composition with Ariel} SKAO's physical and compositional characterization of protoplanetary disks, and its synergy with the ongoing and future ALMA and JWST surveys, will also provide a critical ground-truth for the robust interpretation of the exoplanetary observations of the ESA space mission Ariel \citep{Tinetti2018,Tinetti2022}. Ariel will study what exoplanets are made of, how they formed, and how they evolve, by surveying a diverse sample of about 1000 extrasolar planets, simultaneously in visible and infrared wavelengths. Ariel's observations will provide the first homogeneous survey of the chemical composition of hundreds of transiting exoplanets \citep{Tinetti2018,Zingales2018,Edwards2019} and will allow to systematically explore their origins and the roots of the outstanding exoplanetary diversity \citep{Tinetti2018,Turrini2018,Turrini2022}. The reliable interpretation of Ariel's exoplanetary observations requires disentangling the impact of their formation histories from that of their formation environments \citep{Turrini2021,Turrini2022,Pacetti2022,Dash22}. This, in turn, requires the in-depth understanding of the diversity of planet-forming environments across time and space in protoplanetary disks \citep{Eistrup2016,Eistrup2018,Booth2019,Pacetti2022,Pacetti2025} that the SKAO-ALMA-JWST synergy will provide.

\section{Constraints on the chemical buildup of planets and their atmospheres} 
\label{sect:planets}



The unprecedented sensitivity of SKA to centimeter-wavelength emission from the central regions of disks, close to the midplane, positions the observatory to provide transformative contributions to our understanding of planet formation and exoplanet characterization through the disk-chemistry studies it will enable. 

Planets form by accreting gas and solids from their natal protoplanetary disk. Observations and models consistently highlight that disks are diverse and constantly evolving environments. The final composition of planets and their atmospheres, our window into their nature and formation history, is thus determined by the chemical makeup of the local disk environment at the time and place of planetary growth. 
The chemical inventory available to growing planets is controlled both by the initial partitioning of elements inherited from the parent molecular cloud --- between volatile, semi-refractory (e.g., refractory organics, hydrated and ammoniated minerals), and refractory phases --- and by the subsequent physical and chemical evolution of the disk. Beyond local chemistry, dynamical processes such as viscous gas transport, grain growth,  radial drift of dust, and pebbles redistribute material and reshape radial gradients in temperature, density, and ionization, driving further chemical evolution \citep{Booth2019, Houge2025, Pacetti2025}.

The combined outcome of these processes is a radially stratified and temporally evolving chemical structure in the disk midplane, through which planets form and migrate. As they accrete solids and gas from different regions, planets inherit distinct chemical signatures that can later be traced in their atmospheres \citep[e.g.,][]{Oberg2011, Madhusudhan2019, Turrini2021, Schneider2021, Dash22, Crossfield2023}. By mapping the spatial distribution and abundances of key molecular carriers in the midplane, SKA will deliver critical constraints on the initial chemical conditions of planet formation, enabling a predictive link between disk chemistry and atmospheric composition. For example, tracing carbon chains and cyanopolyynes in the midplane can provide indirect constraints on the fraction of carbon sequestered in semi-volatile refractory organic material. Such measurements inform the partitioning of carbon between gas and solids and the emergence of carbon-rich environments, with direct consequences for the diversity of planetary atmospheres \citep[e.g.,][]{Pacetti2025}.

An equally important frontier is the coupling between dust evolution and gas-phase chemistry. The centimeter-sized grains that SKA will directly probe correspond to the pebble population that dominates radial drift and can strongly enrich the gas phase in volatiles when crossing snowlines. By mapping this pebble population, SKA will not only quantify their impact on gas enrichment but also test whether their contribution alone can account for the observed chemical signatures in the gas. If the measured enrichment deviates from what cm-sized pebbles can supply, additional sources --- such as smaller grains \citep{Pacetti2025} or planetesimals \citep{Tanaka2013, Turrini2019, Eriksson2021} --- must be invoked. Observations restricted to the gas phase or providing only partial coverage of the total dust population cannot resolve this degeneracy. By providing simultaneous access to both pebbles and their chemical impact, SKA will uniquely constrain how dust transport regulates disk chemistry and sets the initial conditions of planet formation.

SKA will also provide important information on the recently proposed process of injection of second-generation dust due to the collisions of planetesimals in the observed protoplanetary disks (\citealt{Turrini2019, Gerbig2019, Bernabo2022, testi2022}, see also chapters by \citealt{Garufi01.2026.SKA,Wu01.2026.SKA}). As massive planets form in protoplanetary disks, they affect their surrounding environment in two ways. In disks whose solid phase is dominated by dust and pebbles, the growing planets that reach the pebble isolation mass \citep{Johnansen2017} will create dust traps that effectively halt the inwards diffusion of dust and pebbles, limiting the enrichment of the disk gas downstream of the traps.

In disks whose solid phase is dominated by planetesimals, the forming planets will excite their surrounding planetesimal disk on eccentric and inclined orbits, causing their orbital paths to intersect and thus triggering a chain of dust-producing collisions. This event will inject new dust and pebbles in regions of the disk that would otherwise be dust-depleted by the effect of the dust traps, thus, potentially restarting the pebble accretion, extending the planet formation temporal window \citep{Turrini2023, Sirono2025}. In the presence of giant planets, planetesimals can reach supersonic velocities that will cause them to experience thermal ablation and high energy collisions, releasing non equilibrium species in the gas \citep{Tanaka2013, Turrini2019, Eriksson2021, Polychroni2025}.
In mixed pebble and planetesimal disks all these effects will be at play. SKA will be able to disentangle their contributions, by providing high precision observations of both gas and pebbles in such disks.

\section{Chemical networks and binding energies} 
\label{sect:chem_be}


\begin{figure}
\begin{center}
\includegraphics[scale=0.4]{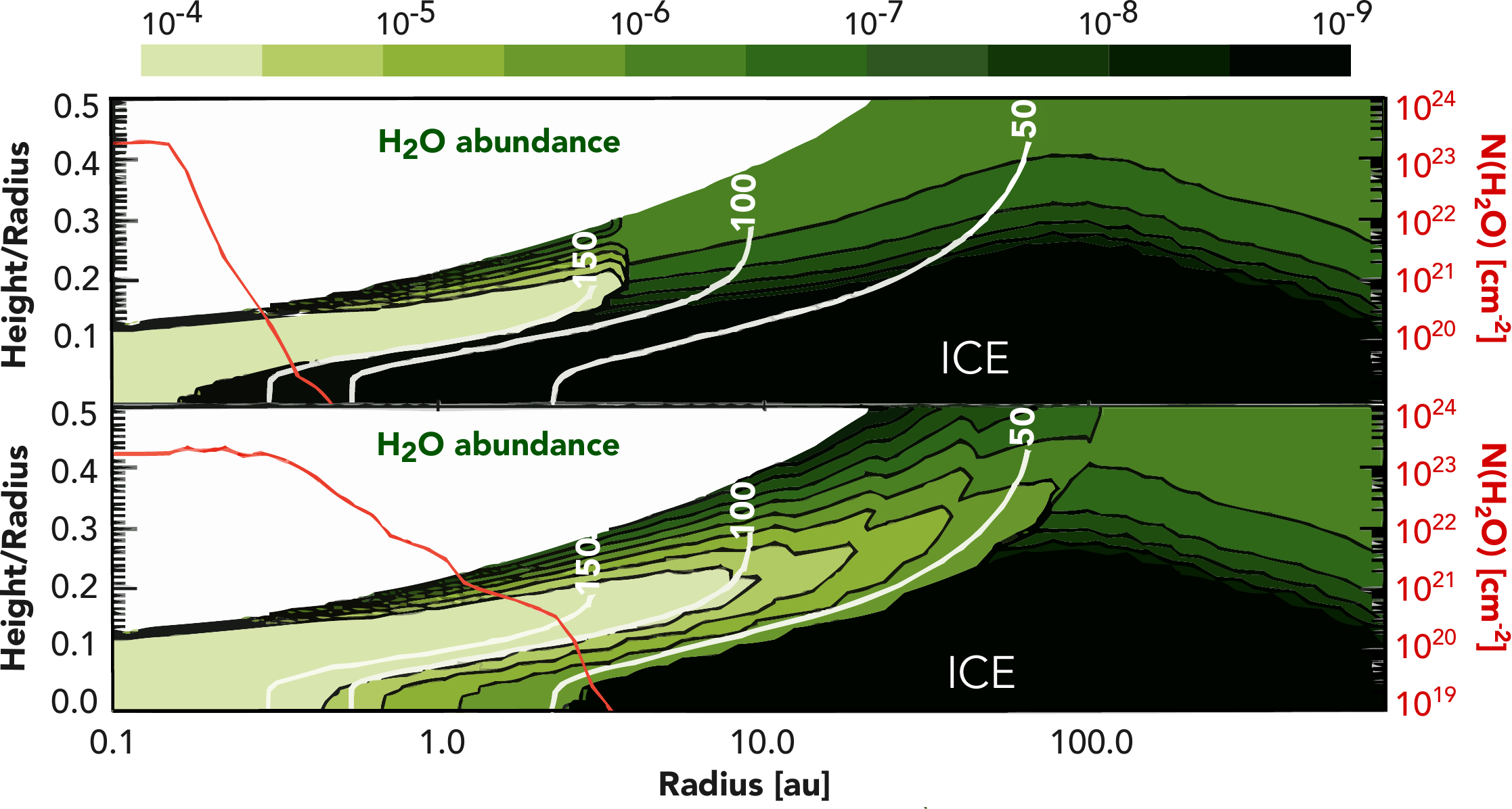}
\caption{Water gas-phase abundance (green colour scale) and column density profile (N(H$_2$O), red solid line) in the protoplanetary disk DM Tau. The figure shows the water abundance obtained using the water BE (5600K) and $v_{\rm des}$ (2$\times$10$^{-12}$ s$^{-1}$) from \citet{wakelam2017} (top panel), and the BE and $v_{\rm des}$ distributions computed in \citet{Tinacci2023-BE} (bottom panel). The physical structure (temperature and density) of the disk of the low-luminosity T Tauri star DM Tau  (0.36 L$\odot$) is taken from \citet{Dominik2005, manara2014}.
The white solid lines mark the dust temperatures of 150, 100, and 50 K. In the black regions, water is frozen onto grain surfaces. Figure adapted from \citet{Tinacci2023-BE}.
}
  \label{fig:tinacci2023}
  \end{center}
\end{figure}

In order to understand the formation and destruction of molecules under interstellar conditions, it is essential to characterize the processes that occur both in the gas phase and on grain surfaces.
Two of the main limitations in our current understanding of astrochemical processes, and hence in the interpretation of astronomical observations, are:
(i) the use of incomplete or outdated reaction networks, both in the gas phase and on grain surfaces;
(ii) the uncertainties in the binding energies (BEs) and pre-exponential factors used to compute desorption rates of species from grain mantles to the gas phase \citep[e.g., ][]{Minissale2022}.

Regarding the first point, many of the reaction rates and products in widely used networks such as the KIDA \citep[Kinetic Database for Astrochemistry;][]{wakelam2024} and UDfA \citep[UMIST Database for Astrochemistry;][]{Umist2022} databases are neither computed using accurate quantum chemical methods nor measured in the laboratory, and therefore need to be carefully evaluated \citep[e.g.,][]{balucani2015formation,skouteris2018genealogical,vazart2020gas,garrod2022,tinacci2023,giani2023revised}. This limitation also applies to more complex chemical networks involving isotopic variants and spin states of major hydrogenated species such as H$_2$, H$_2^+$, H$_3^+$, and their isotopomers \citep[e.g.,][]{Majumdar17}. Since model predictions of molecular abundances are highly sensitive to the adopted network, outdated or incomplete chemistry can lead to unreliable results and complicate the interpretation of molecular formation and destruction pathways.
Revising key reactions, using both quantum chemistry calculations and laboratory experiments, is therefore essential to obtain accurate products and reaction rates.
In view of the future SKA observations, which will allow the detection of many carbon chains, a more robust understanding of their gas-phase formation pathways is urgently needed.
For example, \citet{loison2014} revised the gas-phase chemistry of carbon chains in cold clouds, highlighting the importance of polyynyl radicals (C$_{\rm 2n}$H) for the formation of cyanopolyynes (HC$_{\rm 2n+1}$N). More recently, \citet{Giani2025} revised the formation network of HC$_5$N, showing through combined theoretical and experimental studies that several neutral-neutral reactions play a key role in the production of cyanopolyynes. Despite such improvements, models still fail to reproduce the observed abundances of complex C-bearing species \citep[e.g.,][]{loison2014,2021loomis,giani2025b}. In particular, the observed abundances of the heavier species are especially difficult to reproduce, most likely because of the incompleteness of current chemical networks \citep{giani2025b}.
Thus, the chemistry of polyynyl radicals, cyanopolyynes, and their interconnection remains poorly constrained.

Turning to the second point, recent experiments and theoretical studies have shown that each species is characterized by a distribution of BEs, rather than a single value. This distribution depends on both the adsorption site and the orientation of the species on the grain surface \citep[e.g.,][]{Tinacci2022}.
Similarly, the pre-exponential factor ($v_{\rm des}$), commonly assumed to be 10$^{12}$ s$^{-1}$ for all species, can in fact vary by several orders of magnitude depending on the molecule and the method used for its estimate \citep{ferrero2022}. Such variations can shift desorption temperatures by 20-30 K.
So far, BE and $v_{\rm des}$ distributions have been experimentally or theoretically determined for only a limited number of species \citep[e.g.,][]{collings2004,wakelam2017,das2018,ferrero2020,corazzi2021,Tinacci2022,perrero2024,perrero2024b,bariosco2025}. Yet most astrochemical models still adopt a single BE value per species, an assumption that not only affects desorption rates but also surface diffusion (since diffusion barriers scale with the BE), thereby impacting both the formation efficiency of more complex molecules and their release in gas-phase \citep[e.g., ][]{cuppen2017,Minissale2022,ligterink2025}.
Models of protoplanetary disk chemistry likewise typically assume a single BE value for all species, with only a few exceptions \citep[e.g.,][]{Grassi2020}. 
However, by implementing the BE and $v_{\rm des}$ distributions of water, \citet{Tinacci2023-BE} showed that the frozen-water region becomes smaller by nearly an order of magnitude in radius compared to models using a single BE (5600 K) and  (2$\times$10$^{-12}$ s$^{-1}$) \citep{wakelam2017} (see Fig. \ref{fig:tinacci2023}). 
At the same time, water ice begins to desorb at much larger distances extending the region where gaseous water abundance is $\ge$1$\times$ 10$^{-8}$, both on the disk plane and on the layers above it.
Adopting a BE distribution therefore implies that, instead of a single, well-defined snowline (typically assumed at $\sim$5 au for water), there exists an extended "water transition zone" spanning several astronomical units.
Furthermore, \citet{Boitard2025} showed that, using the BE
distribution of \citet{Tinacci2023-BE} in a model of the protosolar nebula with a temperature profile between 145 and 200 K at 1 au, 
while the bulk of ice is indeed desorbed at larger
distances than 1au, a small fraction (between
$\sim$0.04 and 2.5 wt\%) remains attached to the
grain surfaces inward. This small fraction of frozen water can fully account for the Earth's water content, implying that terrestrial water could be mostly inherited from the dust grains that were in the Earth's orbit, with no necessity of migration of outer ones.
Accurate determinations of BE distributions and $v_{\rm des}$ for a wider range of molecules, together with their correct implementation in astrochemical models, are thus essential for a realistic description of snowlines in protoplanetary disks.
For carbon chains, estimates of the BE distributions currently exist only for HC$_3$N and HC$_5$N on amorphous and crystalline water ice \citep{berta2024}, while data for polyynes and larger cyanopolyynes are still missing. 
Future studies focusing on the BEs and $v_{\rm des}$ of larger carbon chains, along with their incorporation into disk chemical models, will be fundamental to fully exploit and interpret upcoming SKA observations.

\section{Comparing the chemistry of disks with that of pristine objects in the SS}
\label{sect:comets}

The study of small bodies is essential for understanding the processes at the basis of planet formation. In fact, while planets have undergone significant geological processing, differentiation, and atmospheric alteration, small bodies such as comets, asteroids, and trans-Neptunian objects have largely retained their protoplanetary disk compositions and structures. Thus, comparing their chemical composition to that of star- and planet-forming regions can provide important clues into the processes that shaped our and other planetary systems, as well as revealing whether the material in planetary systems is inherited from the early stages of star formation or reprocessed in the disk before being incorporated into planetesimals. 

\begin{figure}
\begin{center}
\includegraphics[scale=0.3]{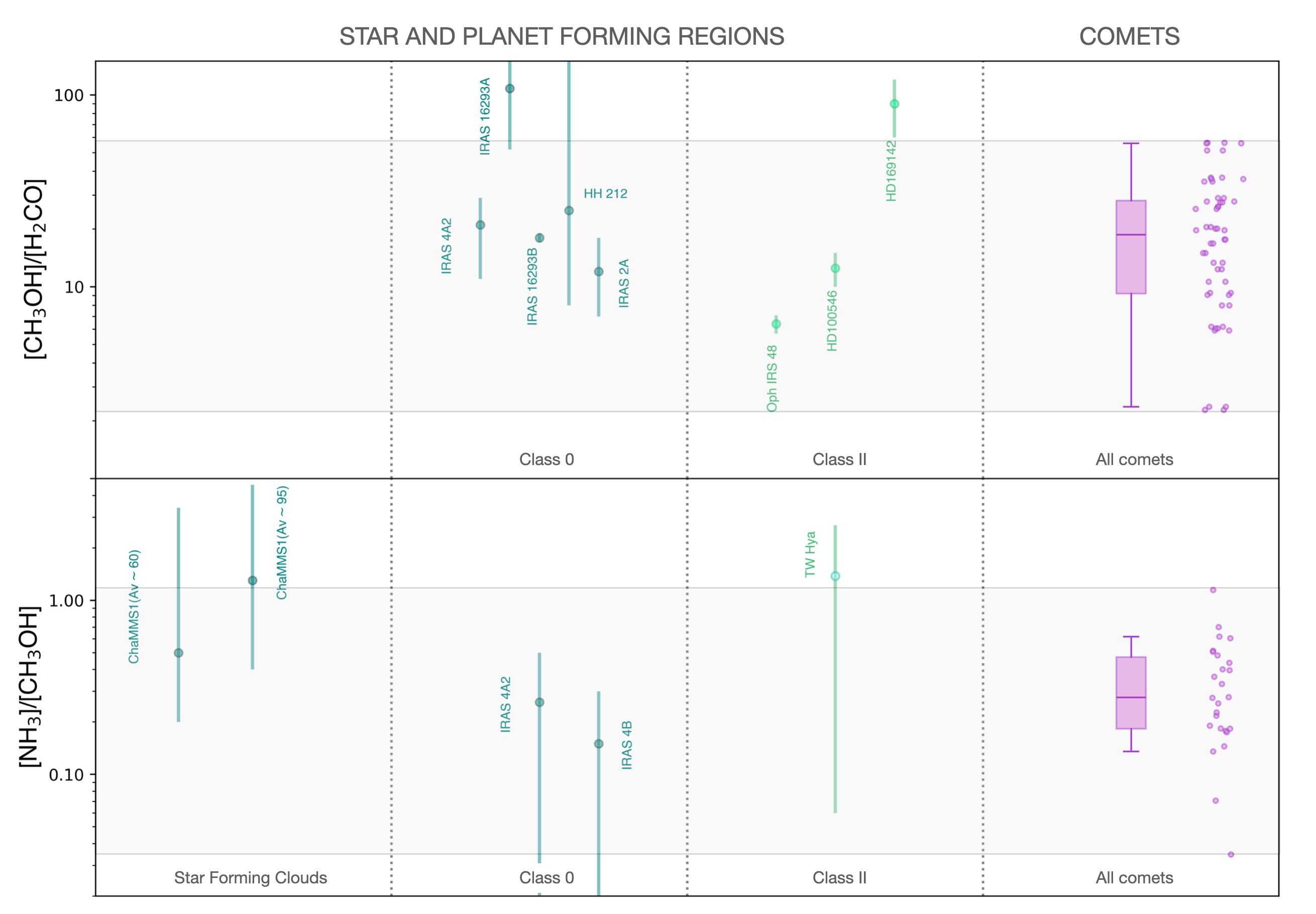}
  \caption{Comparison of the [CH$_3$OH]/[H$_2$CO] and [CH$_3$OH]/[NH$_3$] abundance ratios in star forming clouds, hot corinos around Class 0 protostars, inner regions of Class II disks, and comets. Adapted from \cite{Lippi2024}.} 
  \label{fig:cometsvsdisks}
  \end{center}
\end{figure}

In their recent work, \cite{Lippi2024} implemented the first statistical comparison of the abundances of methanol, formaldehyde, and ammonia and their ratios in a sample of 35 comets and 11 planet-forming systems, i.e., young Sun-like stars with ages ranging from
10$^4$ yr (Class 0 sources) to 10$^7$ yr (Class II disks). The analysis of [CH$_3$OH]/[H$_2$CO] and [CH$_3$OH]/[NH$_3$] abundance ratios showed a
consistency between Class 0 hot-corinos, inner regions of Class II disks, and comets, supporting an inheritance scenario (see Fig. \ref{fig:cometsvsdisks}). To better understand the connections between comets and star- and planet forming regions, as well as how chemical complexity emerges during the different stages of planet formation, these comparative studies must continue and expand, including  also increasingly complex species. 

While the advent of ALMA allowed investigation of small C-chains (eg. c-C$_3$H$_2$, HC$_3$N) as well as complex organic molecules released in the inner sublimation region of disks (T= 80-200 K), such as CH$_3$OH, CH$_3$OCH$_3$, CH$_3$OCHO, CH$_3$CHO, HCOOH \citep[e.g., ][]{Fadul2025a,Booth2024a,Lee2022,Ilee2021}, SKA will be the only facility that will enable study of heavier C-chains (e.g. cyanopolyynes) and rings (e.g. benzonitrile) originating from a colder disk layer close to the midplane, where planets form (see Sect. \ref{sect:C-molecules}). 

Some of these molecules are already investigated in comets, mainly through radio surveys; moreover, they were identified in 67P/Churyumov-Gerasimenko (67P) by the Rosina mass spectrometer onboard of the ESA-Rosetta spacecraft (e.g., HC$_3$N, HCOOH, CH$_3$OCH$_3$; for a complete list of molecules detected in comets see Table 3 in \citet{biver2024}, and reference therein). We anticipate that the identification, quantification, and characterization of heavier C-chains and rings in comets will follow advances in star- and planet-forming regions. These progresses will be supported by new observing facilities (ELT, SKA) and upcoming in-situ missions (e.g., Comet Interceptor).




\section{Summary}
\label{sect:conclusions}

As discussed in several chapters of the Cradle of Life WG, SKAO will open a new window for astrochemical studies at cm wavelengths by exploiting the SKA-Mid Band 5 frequency range ($4.6-15.4$ GHz).
While the exploration of the chemistry of protostellar envelopes, accretion streamers, and outflows on a few hundreds to thousands au scales will be achieved using the initial AA$^{*}$ configuration of SKA-Mid Band 5, the characterization of the faint line emission from disk  on $100$ au scales will require the AA4 final configuration, including the MeerKAT antennas.
The AA4 configuration will make available the largest baselines (up to 160 km), and collecting area, thus allowing to detect line emission from the disk down to a $3\sigma$ sensitivity of $570$ mK
(120 $\mu$Jy/beam) and
$180$ mK (48 $\mu$Jy/beam) with 100 and 1000 hours integration times, respectively (sensitivities obtained with the SKA-Mid sensitivity calculator in Band 5 adopting a spectral resolution of 0.34 km s$^{-1}$ and a beam of $\sim 1"-2"$, i.e. $100-200$ au for nearby disks, which allows avoiding beam dilution and recovering disk-integrated line intensities). 
According to our predicted spectra, these unprecedented high sensitivities will allow detecting line emission from simple organic molecules, such as H$_2$CO and CH$_3$OH, from protostellar and outbursting disks, as well as emission from cyanopolyynes, HC$_3$N and HC$_5$N, in carbon-rich protoplanetary disks with 100 hours of SKA-Mid integration time. 
Integration times of 1000 hours will be needed to detect larger iCOMs (e.g., methyl formate, formamide, glycoladehyde), as well as more complex C-chains (up to HC$_9$N, and C$_8$H).
The extremely high request of integration time will be compensated by the field of view of SKA-Mid ($6.7'$ at 12.5 GHz (Band 5b), and $12.5'$ at 6.7 GHz (Band 5a)), significantly larger compared to that of (sub-)millimeter interferometers. This will allow observing several disks with a single pointing, and to cover the disk population of a nearby star-forming region in a few pointings, thus providing an unprecedented opportunity to sample the disk chemistry on a statistical significant sample of disks.
Based on the science cases explored in this chapter, we foresee that the SKAO will open a window on a new domain of the disk chemistry allowing to search, for the first time, complex carbon-bearing molecules in disks, and to explore the planet formation region which is obscured by dust opacity at mm wavelengths.
Finally, as highligthed in \citet{Codella2015}, and in the SKA Band 6 White Paper\footnote{\url{https://www.skao.int/sites/default/files/documents/d38-ScienceCase_band6_Feb2020.pdf}} the expansion of SKA-Mid to frequencies $>15$ GHz (up to 25 or even 50 GHz), as part of a future Observatory Development Program, will further enhance the capabilities of the SKAO to detect prebiotic molecules such as glycine.

\section*{Acknowledgements}
\textit{
LP, ClCo, and GS acknowledge financial support
under the National Recovery and Resilience Plan (NRRP), Mission 4, Component 2, Investment 1.1, Call for tender No. 104 published on 2.2.2022 by the Italian Ministry of University and Research (MUR), funded by the European Union - NextGenerationEU-Project Title 2022JC2Y93 Chemical Origins: linking the fossil composition of the Solar System with the chemistry of protoplanetary disks - CUP C53D23001600006 - Grant Assignment Decree No.
962 adopted on 30.06.2023 by the Italian Ministry of Ministry of University and Research (MUR).
ClCo, LP, and GS also acknowledge the PRIN-MUR 2020  BEYOND-2p (Astrochemistry beyond the second period elements, Prot. 2020AFB3FX), the project ASI-Astrobiologia 2023 MIGLIORA
(Modeling Chemical Complexity, F83C23000800005), the INAF funding GO 2024 "ICES: Tracking the history of ices from the cradles of planets to comets", GO 2023
"PROTO-SKA (Exploiting ALMA data to study planet forming disks: preparing the advent of SKA, C13C23000770005), Minigrant 2023 TRIESTE ("TRacing the chemIcal hEritage of our originS: from proTostars to planEts"), and Minigrant 2022 "Chemical Origins".
EB acknowledges the support from the Italian Ministry for Universities and Research under the Italian Science Fund (FIS 2 Call - Ministerial Decree No. 1236 of 1 August 2023) and the Next Generation EU funds within the National Recovery and Resilience Plan (PNRR), Mission 4 - Education and Research, Component 2 - From Research to Business (M4C2), Investment Line 3.1 - Strengthening and creation of Research Infrastructures, Project IR0000034 - "STILES - Strengthening the Italian Leadership in ELT and SKA". 
CeCe, ClCo, LP acknowledge the EC H2020 research and innovation
programme for: (i) the project "Astro-Chemical Origins" (ACO, No 811312), and (ii) the European Research Council (ERC) project "The Dawn of Organic Chemistry" (DOC, No 741002).
 LC, IJ-S, and VMR, acknowledge support from grant no. PID2022-136814NB-I00 by MICIU/AEI/10.13039/501100011033 and by ERDF, UE. GE and PRM thank project PID-2022-137980NB-I00 funded by the Spanish Ministry of Science and Innovation / State Agency of Research MCIN / AEI/10.13039/501100011033 and by "ERDF A way of making Europe". 
 EP acknowledges the support from: 
the European Research Council via the Horizon 2020 Framework Programme ERC Synergy "ECOGAL" Project GA-855130
the Italian Space Agency (ASI) through the ASI-INAF grant No.2016-23-H.0 plus addendum No.2016-23-H.2-2021, and the ASI-INAF contract No.2021-5-HH.0.
AC received financial support from the European Research Council (ERC) under the European Union's Horizon 2020 research and innovation programme (ERC Starting Grant "Chemtrip", grant agreement No 949278).
C.W.~acknowledges financial support from the Science and Technology Facilities Council and UK Research and Innovation (grant numbers ST/X001016/1, MR/T040726/1, and MR/Z00019X/1).
JDI acknowledges support from an STFC Ernest Rutherford Fellowship (ST/W004119/1). Y.W. acknowledges the EACOA Fellowship awarded by the East Asia Core Observatories Association. This research was supported by the funding from the National SKA Program of China under Grant No. 2025SKA0120100. C.-F.L. acknowledges support from the National Science and Technology Council (NSTC) of Taiwan (112-2112-M-001-039-MY3).
I.J.-S. acknowledges support from the ERC Consolidator Grant "OPENS" (grant agreement N. 101125858) funded by the European Union. JEP was supported by the Max-Planck Society. Part of this research by M.N. was carried out at the Jet Propulsion
Laboratory, California Institute of Technology, under a contract with the National Aeronautics and Space Administration (80NM0018D0004).}

\bibliographystyle{abbrvnat-maxbibnames4}
\bibliography{chapter} 

\end{document}